\documentclass[twocolumn]{aastex63}

\usepackage[version=4]{mhchem}
\usepackage{amsmath,amssymb}
\usepackage{braket} 
\usepackage{etoolbox}
\usepackage{lineno}
\received{}
\revised{}
\accepted{}
\submitjournal{ApJL}

\shorttitle{High water D/H ratio of 3I/ATLAS is consistent with a low-metallicity origin}
\shortauthors{Furuya et al.}

\begin{document}

\title{High water D/H ratio of the interstellar object 3I/ATLAS is consistent with a low-metallicity origin}

\correspondingauthor{Kenji Furuya}
\email{kenji.furuya@riken.jp}

\author[0000-0002-2026-8157]{Kenji Furuya}
\affiliation{RIKEN Pioneering Research Institute, 2-1 Hirosawa, Wako-shi, Saitama 351-0198, Japan}

\author[0000-0001-8233-2436]{Martin Cordiner}
\affiliation{Solar System Exploration Division, NASA Goddard Space Flight
Center, 8800 Greenbelt Rd, Greenbelt, MD 20771, USA}
\affiliation{Department of Physics, Catholic University of America, 620 Michigan
Ave. NE, Washington, DC 20064, USA}

\author[0000-0002-8130-0974]{Dominique Bockel{\'e}e-Morvan}
\affiliation{LIRA, Observatoire de Paris, Universit\'e PSL, CNRS, Sorbonne Universit\'e, Universit\'e Paris Cit\'e, CY Cergy Paris Universit\'e, 5 place Jules Janssen, F-92190 Meudon, France}

\author[0000-0002-2668-7248]{Dennis Bodewits}
\affiliation{Department of Physics, Auburn University, Edmund C. Leach Science Center, 36382,  Auburn, AL, USA}

\author[0000-0001-7335-1715]{Colin Orion Chandler}
\affiliation{Dept. of Astronomy \& the DiRAC Institute, University of Washington, 3910 15th Ave NE, Seattle, WA 98195, USA}
\affiliation{LSST Interdisciplinary Network for Collaboration and Computing, 933 N. Cherry Avenue, Tucson, AZ 85721, USA}
\affiliation{Dept. of Astronomy \& Planetary Science, Northern Arizona University, PO Box 6010, Flagstaff, AZ 86011, USA}

\author[0000-0001-7479-4948]{Maria N. Drozdovskaya}
\affiliation{Department of Chemistry, Biochemistry and Pharmaceutical Sciences (DCBP), Universit{\"a}t Bern
Freiestrasse 3, 3012 Bern, Switzerland}

\author[0000-0002-6006-9574]{Nathan X. Roth}
\affiliation{Solar System Exploration Division, NASA Goddard Space Flight
Center, 8800 Greenbelt Rd, Greenbelt, MD 20771, USA}
\affiliation{Department of Physics, American University, 4400 Massachusetts Avenue NW, Washington, DC 20016, USA}

\author[0000-0002-2662-5776]{Geronimo Villanueva}
\affiliation{Solar System Exploration Division, NASA Goddard Space Flight
Center, 8800 Greenbelt Rd, Greenbelt, MD 20771, USA}



\begin{abstract}
Recent JWST observations have revealed unusually high $^{12}$C/$^{13}$C ratios in carbon-bearing molecules of the interstellar object 3I/ATLAS, consistent with formation in a lower-metallicity environment than the present-day local interstellar medium (ISM).
3I/ATLAS also exhibits an exceptionally high water D/H ratio, exceeding those in Solar System comets and nearby low-mass star-forming regions. 
Here we investigate whether this high water D/H ratio can be reproduced in a low-metallicity formation scenario, using gas-ice astrochemical models.
Assuming that the water observed in 3I/ATLAS was inherited from the parent molecular cloud and core, we perform a grid of astrochemical models covering the cloud to core stages, varying the gas density, ultraviolet radiation field ($\chi$), cosmic-ray ionization rate ($\zeta$), and metallicity, while solving thermal balance for the gas temperature. 
We find that lower metallicity enhances \ce{H3+} deuteration and, more importantly, its transfer to water ice.
In contrast, water D/H ratio depends non-monotonically on $\chi$ and $\zeta$, because of competing chemical and thermal effects.
In our models, the observed water D/H ratio is most readily reproduced at subsolar metallicities, $\lesssim0.5Z_\odot$, and relatively high cloud densities of $\sim$10$^4$ cm$^{-3}$ without strong constraints on either $\chi$ or $\zeta$, as long as $\zeta<10^{-15}$ s$^{-1}$.
The D/H ratio of methane normalized by that of water is not sensitive to the metallicity, being consistent with the similar values observed in 67P/Churyumov–Gerasimenko and 3I/ATLAS.
These results suggest that water deuteration may provide a complementary probe of the metallicity and physical condition of the parent molecular cloud and dense core of interstellar objects.
\end{abstract}





\section{Introduction}
The isotopic compositions of cometary volatiles provide unique probes of the physical and chemical environments in which comets form \citep[e.g.,][]{bockelee-Morvan15,nomura23}.
Recent JWST observations of the interstellar comet 3I/ATLAS revealed unusually high $^{12}$C/$^{13}$C ratios in CO and \ce{CO2}, the major volatile carbon reservoirs \citep[123--191;][]{cordiner26}.
These ratios are much higher than the elemental $^{12}$C/$^{13}$C ratio in the local ISM \citep[68 $\pm$  15;][]{milam05}, the Sun \citep{lyons18}, and the molecular $^{12}$C/$^{13}$C ratios in the Solar System comets \citep[$\sim$90;][]{bockelee-Morvan15}.
The $^{12}$C/$^{13}$C ratio in CN is also found to be high \citep[$>$100;][]{opitom26}.
These measurements suggest that the bulk volatile carbon in 3I/ATLAS is depleted in $^{13}$C relative to the local ISM and the Solar System, a trend predicted by Galactic chemical evolution models for old and relatively metal-poor Galactic environments \citep[$Z <1$, where $Z$ is the metallicity relative to solar metallicity, $Z_\odot$;][]{cordiner26}.
This interpretation is also compatible with the comet's unusually high heliocentric velocity and incoming trajectory, which suggests an origin in an old stellar population with subsolar metallicity \citep{hopkins25,taylor25}.

Another isotopic feature of 3I/ATLAS is its exceptionally high water D/H ratio, $(9.8\pm 0.6) \times10^{-3}$ \citep[\citealt{cordiner26}; see also][]{SalazarManzano26}.
This value is much higher than those observed in the low-mass protostellar sources ($\sim$10$^{-3}$) and the Solar System comets \citep[$\sim$3$\times$10$^{-4}$;][and references therein]{nomura23}.
A substantial fraction of water in comets may be inherited from the parent molecular cloud and core, as suggested by the similar water D/H ratios in low-mass protostars and Solar System comets \citep[e.g.,][]{nomura23}.
Models suggest that the water D/H ratio established during the cloud and core stages is largely preserved in protoplanetary disks, although some processing can occur.
In Class 0/I disks, high-temperature gas-phase isotope-exchange reactions between HDO and \ce{H2} can reduce the water D/H ratio in the innermost disk regions ($\gtrsim500$ K), 
with the processed water subsequently transported to outer disk regions through turbulent mixing \citep[e.g.,][]{yang13,owen15}.
In Class II disks, the deuteration of water ice is inefficient because of the limited availability of free oxygen for water-ice formation and weak ionization near the disk midplane \citep{cleeves14}.
Thus, disk processing is expected to preserve or reduce the inherited water D/H ratio.
The exceptionally high water D/H ratio of 3I/ATLAS may therefore retain information on conditions in its natal molecular cloud and core environment and thus provide a complementary constraint on the proposed low-metallicity origin.

Observational constraints on the water D/H ratio in low-metallicity star-forming regions remain sparse.
HDO has been detected towards high-mass star-forming regions in the extreme outer Galaxy \citep[$\lesssim 0.25Z_\odot$;][]{shimonishi21} and the Large Magellanic Cloud \citep[$\sim (1/3-1/2)Z_\odot$;][]{sewilo22}, but the water D/H ratio has not been constrained yet.

In this Letter, we investigate how the metallicity affects the water D/H ratio in molecular clouds and cores using gas--ice astrochemical models. 
We adopt the working hypothesis that the water D/H ratio observed in 3I/ATLAS was inherited from the parent molecular cloud and core, and was delivered to the protoplanetary disk and subsequently incorporated into 3I/ATLAS without significant chemical reprocessing.
Under this assumption, we test whether low-metallicity conditions can reproduce the extreme water deuterium enrichment observed in 3I/ATLAS.

Lower metallicity can affect deuterium fractionation through several chemical effects.
It is well established that the main driver of the deuterium fractionation is the following reaction \citep{watson76}:
\begin{equation}
\ce{H3+} + \ce{HD} \rightleftharpoons \ce{H2D+} + \ce{H2} + \Delta E. \label{react1}
\end{equation}
At low temperatures, the backward reaction is suppressed by its endothermicity, allowing \ce{H2D+} to become enhanced relative to \ce{H3+}.
The efficiency of this fractionation is reduced by ortho-\ce{H2}, whose internal energy can promote the backward reaction \citep[e.g.,][]{pagani92}.
The deuterium enrichment in \ce{H3+} isotopologues is subsequently transferred to other gaseous molecules and to icy molecules.
In lower metallicity environments, heavy-element-bearing species tend to be less abundant, reducing the destruction rate of \ce{H3+} isotopologues.
This directly enhances deuterium fractionation \citep[e.g.,][]{dalgarno84} and also increases the conversion rate of ortho-\ce{H2} to para-\ce{H2} through proton-exchange reactions with \ce{H3+} (and with \ce{H+}), which further promotes deuterium fractionation \citep[e.g.,][]{gerlich02}.
However, these effects alone do not determine the water D/H ratio, because metallicity also alters the coupled gas-ice chemistry that regulates the deuterium fractionation of water ice as we discuss below.




\section{Model Setup} \label{sec:modelsetup}
To explain the observed HDO/\ce{H2O} 
\footnote{
The D/H ratio of water corresponds to the ratio HDO/(\ce{H2O} + HDO) $\approx$ 0.5 $\times$ HDO/\ce{H2O}, where the approximation holds when the amount of HDO is negligible compared to \ce{H2O}.
}
and \ce{D2O}/HDO column density ratios in the warm ($\gtrsim$100 K) inner envelopes of nearby low-mass protostars, where water ice has sublimated, \citet{furuya16} proposed a two-stage scenario for water ice formation in the prestellar phase.
First, in a molecular cloud where interstellar ultraviolet (UV) radiation is not fully shielded, most water ice is formed without significant deuterium fractionation.
In the subsequent denser core, the UV radiation is fully shielded, CO starts to freeze out and the \ce{H2} ortho-to-para ratio (OPR) decreases.
There, an additional but smaller amount of water ice is formed with significant deuterium fractionation.

A key feature of this scenario is that most water ice is produced during the molecular cloud stage, whereas the deuterium-enriched component is added during the core stage. 
This is consistent with observational evidence that \ce{H2O} ice forms at lower visual extinctions than CO ice and \ce{CH3OH} ice \citep{whittet13,boogert15} and that the D/H of methanol is higher than that of water in nearby low-mass protostars \citep[e.g.,][]{jorgensen18,zeng25}.
\citet{furuya16} assumed the standard interstellar UV radiation field, the cosmic-ray ionization rate of 10$^{-17}$ s$^{-1}$, and the elemental abundance adequate for the local ISM.
In this work, following this two-stage scenario, we investigate how the water D/H ratio depends on the physical conditions of the prestellar phase, with particular emphasis on metallicity.

\subsection{Physical model}
\citet{furuya15} and \citet{furuya16} studied the formation and deuteration of water ice, using one-dimensional hydrodynamical models of molecular cloud formation and the collapse of dense cores.
Here, rather than employing time-dependent physical models, we adopt a simplified two-step toy model.
Specifically, we perform two sequential pseudo-time-dependent calculations representing the molecular cloud and denser core stages, under fixed physical conditions in each stage.

\subsubsection{Cloud stage} \label{sec:cloudphys}
It is thought that water ice mantles start to accumulate on dust grains when the formation and photodesorption rates of water ice are balanced \citep{tielens05, hollenbach09}.
This is supported by the three-dimensional magneto-hydrodynamical simulations of the molecular cloud formation with post-processing gas-ice chemical network simulations \citep{komichi26}.
For a given UV flux, the critical visual extinction is given by
\begin{equation}
A_{V,\ {\rm crit}} = \frac{1}{\gamma} \ln \left[ \frac{\chi FY_{\rm phdes}}{(1-\alpha) x_{\rm O}n_{\rm cl}v_{\rm th,\,O}} \right], \label{eq:av_crit}
\end{equation}
where $Y_{\rm phdes}$ is the photodesorption yield of water ice per incident FUV photon, $x_{\rm O}$ is the elemental abundance of oxygen, $n_{\rm cl}$ is the cloud gas density, $v_{\rm th,\,O}$ is the thermal velocity of atomic oxygen, and $\gamma$ accounts for the different attenuation of visible and FUV radiation \citep[e.g.,][]{vandishoeck06}.
$\chi = 1$ corresponds to the Habing field \citep{habing68} with the FUV flux of $F = 10^8$ photons cm$^{-2}$ s$^{-1}$.
Since $A_{V,\rm crit}$ depends on the gas temperature ($T_g$) through $v_{\rm th}$, while $T_g$ itself depends on $A_{V,\rm crit}$, the two quantities are solved iteratively. $T_g$ is obtained by balancing heating (photoelectric emission and cosmic rays) and cooling (\ce{C+}, O, and CO line emission) using the gas-phase chemical network, neglecting grain surface chemistry for simplicity.
Once $A_{V,\ {\rm crit}}$ is determined, dust temperature is estimated by using Eq. 8 of \citet{hocuk17},
\begin{equation}
T_d = [11+5.7\tanh(0.61-\log_{10}(A_{V,\ {\rm crit}}))](\chi/1.7)^{1/5.9},
\end{equation}
adopting a floor value of 8 K.
The factor of $1/1.7$ converts the radiation-field strength from the Draine field \citep{draine78} adopted by \citet{hocuk17} to the Habing-field units used in this work.
The parameter $\alpha$ in Eq. \ref{eq:av_crit} determines the duration of the cloud stage. 
If water ice forms on the freeze-out timescale of atomic oxygen \citep[$\tau_{\rm freeze}$;][]{hollenbach09}, 
the fraction of water ice that has already formed by time $t$ is $1-\exp(-t/\tau_{\rm freeze})$, where \begin{equation}
\tau_{\rm freeze} \approx 4 \times 10^5 (Z/1)^{-1} (n_{\rm cl}/10^4 \,\, {\rm cm}^{-3})^{-1} \,\, {\rm yr}.
\end{equation}
Here we assumed that the dust-to-gas mass ratio scales with metallicity and adopted a fixed, uniform dust radius of 0.1 $\mu$m.
Defining $\alpha$ as the nominal fraction of water ice formed during the cloud stage under this simplified freeze-out approximation, we set the duration of the cloud stage to $-\tau_{\rm freeze}\ln(1-\alpha)$.
The water ice formation scenario by \citet{furuya16} suggests that most water ice forms during the cloud stage, corresponding to $\alpha \sim 1$.
For $\alpha$ = 0.90 and 0.95, the corresponding cloud-stage duration is $\sim$2.3$\tau_{\rm freeze}$ and $\sim$3$\tau_{\rm freeze}$, respectively.

\subsubsection{Core stage}
The final abundance at the cloud stage is used for the initial abundance of the subsequent core stage.
The gas density is assumed to be $10n_{\rm cl}$, and the visual extinction is fixed at $A_V=10$ mag, for which the external UV radiation field is effectively negligible.
The value of $\zeta$ is set to be the same as in the cloud stage.
As for the cloud stage, $T_g$ is obtained by solving the balance between heating and cooling, while the dust temperature is calculated using Eq. 8 of \citet{hocuk17}, with the floor value of 8 K.
The lifetimes of starless cores have been observationally estimated from the number of starless cores relative to young stellar objects \citep[e.g.,][]{Ward-Thompson07,takemura23,moon25}.
Recent studies suggest the lifetimes of 5-10 free-fall times ($\tau_{\rm ff}$) for the core density of 10$^4$-10$^5$ cm$^{-3}$ or even longer.
In this work, we assume that the duration of the core stage is $\sim$5$\tau_{\rm ff}$, corresponding to $2\times10^6$ yr and $7\times10^5$ yr for the core densities of 10$^4$ and 10$^5$ cm$^{-3}$, respectively.
Even in models with the core density of 10$^4$ cm$^{-3}$ and $Z=0.2$ (i.e., the longest $\tau_{\rm freeze}$ case among our models), the duration of the core is comparable to $\sim\tau_{\rm freeze}$, allowing further freeze-out of oxygen (i.e., the formation of deuterium-enriched water ice) to proceed during the core stage.

\subsection{Chemical reaction network} \label{sec:network}
We adopt the same gas-ice astrochemical model as in \citet{furuya26}, which includes the non-deuterated and singly, doubly, and triply deuterated forms of all species in our chemical network, as well as the nuclear spin states of \ce{H2}, \ce{H3+}, and their isotopologues.
The model includes gas-phase reactions, interactions between gas and (icy) grain surfaces, and surface reactions.
The gas and ice chemistry is described by the three-phase model \citep{hasegawa93}, assuming that the top four monolayers are chemically active; the rest of the ice mantles is assumed to be inert.
Following \citet{furuya24}, we assume Gaussian binding energy distributions for all surface species. 
Deuterated species are assumed to have the same mean and standard deviation as their non-deuterated counterparts, except atomic D, whose mean binding energy is 21 K higher than that of H (371 K versus 350 K; \citealt{caselli02}). 

For the nuclear spin chemistry of \ce{H2}, all three relevant processes are included; the \ce{H2} formation on grains \citep{watanabe10}, the spin conversion in the gas phase \citep{honvault11,honvault12} and on grain surfaces \citep{furuya19,furuya26}.

Photodissociation and photodesorption rates of water ice isotopologues are calculated in the same way as in \citet{furuya15}, using the results of molecular dynamics simulations by \citet{arasa15}.

\subsection{Parameters} \label{sec:parameters}
The unusually high $^{12}$C/$^{13}$C ratios observed in 3I/ATLAS have been interpreted as signatures of formation in an environment shaped by early Galactic chemical evolution and characterized by a lower metallicity than that of the present-day local ISM \citep{cordiner26}.
Such environments may also differ from nearby molecular clouds in other physical properties. 
Therefore, while metallicity is the primary parameter explored in this work, we also vary the cloud gas density ($n_{\rm cl}$), unattenuated external UV radiation field ($\chi$), and cosmic-ray ionization rate of \ce{H2} ($\zeta$) to isolate the role of metallicity in establishing the water D/H ratio.
Our model has two additional free parameters: $\alpha$ (Sec. \ref{sec:cloudphys}) and the initial value of the \ce{H2} OPR.
These parameters are summarized in Table \ref{table:parameter}.
It should be noted that the effective UV field relevant for photochemistry and gas heating in the cloud phase is largely independent of $\chi$.
This is because $A_{V,\ {\rm crit}}$ is determined by the balance between water ice photodesorption and atomic oxygen freeze-out.
Substituting Eq. \ref{eq:av_crit} into photoreaction rates, which are generally given as $\chi k_1 \exp(-k_2 A_{V,\ {\rm crit}})$, shows that the dependence on $\chi$ largely cancels out, where $k_1$ and $k_2$ are process-dependent parameters.
Instead, the value of $\chi$ affects the dust temperature.

Our fiducial elemental abundances that are available for gas and ice chemistry at $Z=1$ are taken from \citet{aikawa99} and are set to H:He:C:N:O:S:Si:Fe:Na:Mg = 1.00:9.75(-2):7.86(-5):2.47(-5):1.80(-4):9.14(-8):9.74(-9):2.74(-9):2.25(-9):1.09(-8), where $a(-b)$ indicates $a\times10^{-b}$.
For lower metallicity models, the abundances of all elements heavier than He are scaled by a factor of $Z$.
The dust-to-gas mass ratio is also scaled by the same factor.
Initially, hydrogen and deuterium are assumed to be in \ce{H2} and HD, respectively.
The other elements are initially either neutral atoms or atomic ions, depending on their ionization energy.
The elemental D/H ratio is fixed to $1.5\times10^{-5}$ \citep{linsky03}, independent of metallicity.
The uncertainty in the assumed elemental D/H ratio and its impact on the water D/H ratio is discussed in Appendix \ref{app:elemental_dh}.

Observations of nearby galaxies indicate that the dust-to-gas mass ratio ($d/g$) decreases with decreasing metallicity and is approximately proportional to metallicity at $Z \gtrsim 0.2$ \citep[][]{remy-ruyer14,galliano18}.
A steeper dependence ($d/g \propto Z^{1.4^{+1.0}_{-0.3}}$) is also reported within the Galaxy \citep{giannetti17}.
On the other hand, some observations and models of nearby galaxies suggest that the relative contribution of small grains to total dust mass may increase with decreasing metallicity at $Z \gtrsim 0.3$ \citep{relano22}.
This trend could increase the total dust surface area per unit dust mass, partly compensating for the decrease in dust abundance.
In this work, we assume that $d/g = 0.01Z$ with a fixed, uniform dust radius of 0.1 $\mu$m.
These assumptions result in a linear scaling of the total dust surface area per H nucleus with metallicity.

\begin{table}[ht!]
\caption{Summary of adopted parameters}                 
\label{table:parameter}    
\centering                        
\begin{tabular}{c c}      
\hline\hline               
Parameters & Values  \\         
\hline                      
  $Z (Z_\odot)$  & 0.2, 0.5, 1  \\    
   $\chi$ & 0.1, $\mathbf{1}$, 10, 100 \\ 
   $\zeta$ (s$^{-1}$) & 10$^{-18}$, $\mathbf{10^{-17}}$, 10$^{-16}$, 10$^{-15}$  \\
   $n_{\rm cl}$ (cm$^{-3}$) & 10$^{3}$, $\mathbf{10^{4}}$  \\
    $\alpha$ & 0.9, $\mathbf{0.95}$  \\
 Initial \ce{H2} OPR & 10$^{-3}$, $\mathbf{10^{-2}}$, 0.1  \\
\hline                                  
\end{tabular}
\tablecomments{Bold values mark the values adopted in the reference model.
}
\end{table}

\section{Results} \label{sec:result}
\subsection{Metallicity dependence of water D/H ratio}
\begin{figure*}[ht!]
\includegraphics[width=\linewidth]{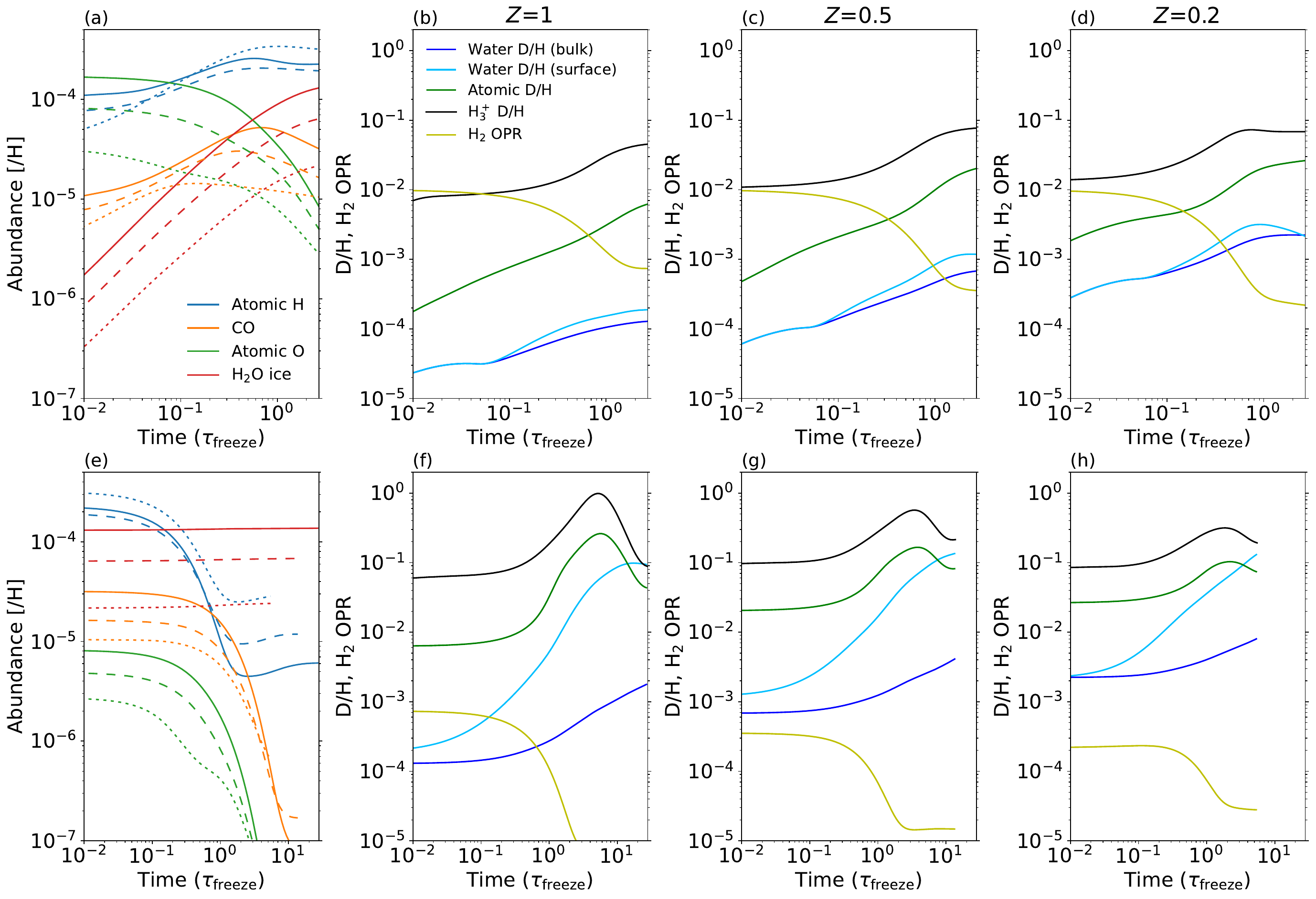}
\caption{
Temporal evolution of selected species and D/H ratios during the cloud (top panels) and core stages (lower panels). Panels (a) and (e) show the abundances of selected species during the cloud and core stages, respectively. Solid, dashed, and dotted lines denote models with $Z = 1$, 0.5, and 0.2, respectively. The remaining panels show the temporal evolution of the D/H ratios and the \ce{H2} OPR for $Z = 1$ (b, f), $Z = 0.5$ (c, g), and $Z = 0.2$ (d, h). In all panels, time is normalized by $\tau_{\rm freeze}$ for the corresponding density and metallicity.
}
\label{fig:time_evol}
\end{figure*}

Figure \ref{fig:time_evol} shows the temporal evolution of the abundances of selected species and the D/H ratios in the cloud and core stages for three metallicities, $Z=1$, 0.5, and 0.2.
The time axis is normalized by $\tau_{\rm freeze}$ to make it easier to compare models with different $Z$.
In all cases shown in the figure, the other parameters are set to the fiducial values listed in Table \ref{table:parameter}.
Note that, unless otherwise specified, the water abundance and D/H ratio discussed in this section refer to values for the entire ice mantle.
We define the D/H ratio of \ce{H3+} as 
\begin{equation}
\frac{x(\ce{H2D+})+2x(\ce{D2H+})+3x(\ce{D3+})} {3x(\ce{H3+})+2x(\ce{H2D+})+x(\ce{D2H+})},
\end{equation}
where $x(i)$ is the abundance of species $i$.
We include \ce{D2H+} and \ce{D3+} because their abundances are not negligible relative to that of \ce{H2D+} in our models. 
Their dissociative recombinations with electrons produce a non-negligible amount of atomic D, which contributes to the formation of HDO ice.

Regardless of $Z$, water ice forms through grain-surface reactions O + H $\rightarrow$ OH, followed by OH + \ce{H2} $\rightarrow$ \ce{H2O} + H in our models.
Although the latter reaction has an activation energy barrier, it is more efficient than OH + H $\rightarrow$ \ce{H2O}, because of quantum tunneling and higher abundance of \ce{H2} than atomic H \citep[e.g.,][]{cuppen07,oba12}.
For $Z=1$, the water D/H ratio reaches $\sim$10$^{-4}$ and $\sim$10$^{-3}$ at the end of cloud and core stages, respectively.
These values are consistent with those obtained by \citet{furuya15} and \citet{furuya16}, despite the simplified treatment of the physical evolution adopted in this work.

The water D/H ratios at the end of the cloud and core stages are higher at lower $Z$. 
The higher water D/H ratios at lower $Z$ arise from two coupled effects: (i) the D/H ratio of \ce{H3+} becomes higher, and (ii) the enhanced D/H ratio of \ce{H3+} is more directly reflected in the gas-phase atomic D/H ratio and subsequently in the water D/H ratio through grain-surface reactions.
Lower $Z$ leads to a lower CO abundance, which reduces the destruction rate of \ce{H3+} isotopologues, and to a lower \ce{H2} OPR, which suppresses the backward reaction of Reaction \ref{react1} as the internal energy of ortho-\ce{H2} helps overcome the endothermicity of the backward reaction.
Together with longer freeze-out timescale, these effects result in a higher D/H ratio of \ce{H3+} at a given normalized time $t/\tau_{\rm freeze}$.
Note that at a fixed $t/\tau_{\rm freeze}$, the lower-$Z$ models correspond to a longer physical time. 
However, the D/H of \ce{H3+} already exceeds 10$^{-2}$ in the cloud stage regardless of $Z$, and this ratio alone does not determine the water D/H ratio. 
More importantly, the atomic D/H ratio follows the D/H ratio of \ce{H3+} more closely at lower $Z$, leading to more efficient deuterium enrichment of water ice on grain surfaces.

In the cloud stage, surface water ice undergoes repeated cycles of photodissociation and reformation.
This is because water ice formation and photodesorption are approximately balanced at $A_{\rm V,crit}$ by definition, 
and because photodissociation is more efficient than photodesorption \citep{arasa15}.
The main branch of \ce{H2O} ice photodissociation produces OH and H, with the latter being released into the gas phase \citep{arasa15}.
The release of atomic H into the gas phase lowers the atomic D/H ratio below the D/H ratio of \ce{H3+}.
This effect is more important at higher $Z$ (see Appendix \ref{app:atomicH} for reasoning), suppressing the deuterium enrichment of water ice at higher $Z$.
For these reasons, the water D/H ratio at the end of the cloud stage is higher at lower $Z$.

Note that as discussed in detail by \citet{furuya15}, repeated photodissociation and reformation of water ice itself suppresses HDO enrichment when the surface reaction OH + \ce{H2} $\rightarrow$ \ce{H2O} + H dominates over OH + H $\rightarrow$ \ce{H2O} \citep[see also][]{kalvans17}.
That is, after HDO is photodissociated, the OH fragments preferentially reform \ce{H2O} through reactions with abundant \ce{H2} rather than reforming HDO, efficiently removing deuterium from the water-ice network.
This mechanism is at work regardless of $Z$, making the D/H ratio of surface water ice much lower than the atomic D/H ratio in the gas phase (compare the light-blue and green lines in the figure).

In the core stage, where external UV radiation is well shielded, the D/H ratio of \ce{H3+} is $\sim$0.1 in all fiducial models, which is more directly reflected in the atomic D/H ratio, and highly deuterated water ice forms more readily than in the cloud stage.
However, because the bulk of water ice formation is already completed in the cloud stage in our models, the enhancement of the D/H ratio of bulk water ice is limited (compare the light-blue and darker-blue lines in the figure), 
and the metallicity dependence of the water D/H ratio established during the cloud stage is largely preserved until the end of the core stage.

\begin{figure*}[ht!]
\includegraphics[width=\linewidth]{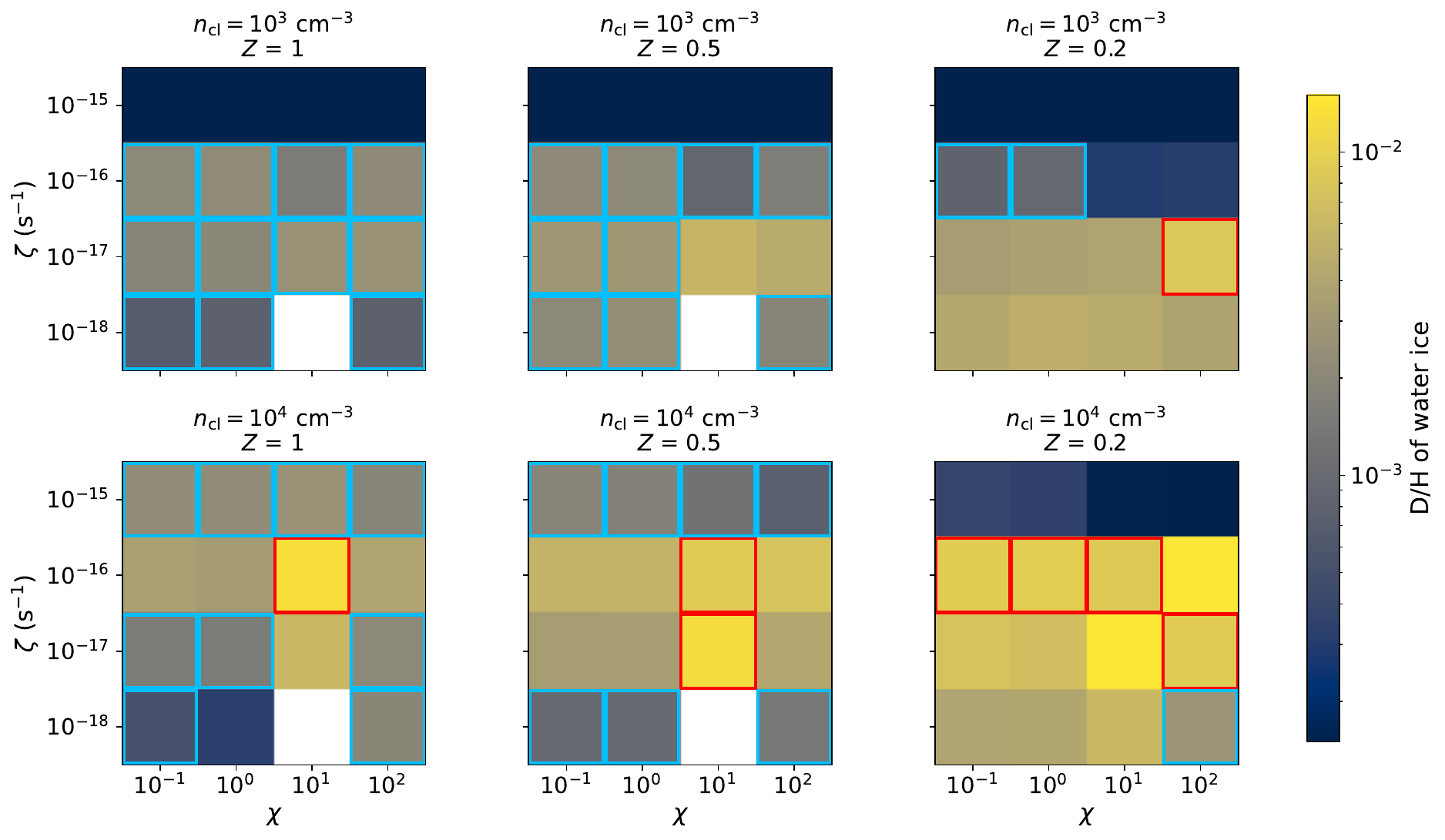}
\caption{
Parameter dependence of the final water D/H ratio (surface+mantle) as a function of the cosmic-ray ionization rate ($\zeta$) and external UV radiation-field strength ($\chi$). The top and bottom panels correspond to cloud densities of $n_{\rm cl}=10^3\ {\rm cm}^{-3}$ and $10^4\ {\rm cm}^{-3}$, respectively. From left to right, the metallicity decreases from $Z=1$ to 0.2. The initial \ce{H2} OPR and the parameter $\alpha$ are fixed at $10^{-2}$ and 0.95, respectively, in all models. Red outlines indicate parameter sets with water D/H ratios within 30 \% of 0.01, and are therefore consistent with 3I/ATLAS within the adopted tolerance. Blue outlines indicate parameter sets with water D/H ratios between $5\times10^{-4}$ and $3\times10^{-3}$, consistent with those observed in nearby low-mass protostars.
White regions indicate parameter sets for which the chemical network calculations did not converge.
}
\label{fig:HDO_H2O__h2opr1e-2_alpha005}
\end{figure*}

\subsection{Grid of models}
Figure \ref{fig:HDO_H2O__h2opr1e-2_alpha005} shows the water D/H ratio at the end of the core stage obtained from a subset of our model grid, fixing $\alpha=0.95$ and the initial \ce{H2} OPR of $10^{-2}$.
As a general trend, a lower $Z$ leads to a higher water D/H ratio.
An exception occurs at $\zeta=10^{-15}$ s$^{-1}$: the water D/H ratio is lower than in the lower-$\zeta$ cases and decreases with decreasing $Z$.
This is due to cosmic-ray heating of the gas, which starts to raise the gas temperature above 20 K when $\zeta=10^{-15}$ s$^{-1}$ and suppresses the deuterium fractionation triggered by Reaction \ref{react1}.
Because the abundances of coolant species (atomic O, \ce{C+}, and CO) are smaller at lower $Z$, the gas temperature is higher at lower $Z$, leading to a lower water D/H ratio.

The water D/H ratio tends to increase with increasing $n_{\rm cl}$, while the dependence on $\chi$ and $\zeta$ is non-monotonic, reflecting the competition among several chemical and thermal effects. 
The dependence of the water D/H ratio on the elemental D/H ratio, $n_{\rm cl}$, $\zeta$, and $\chi$ is discussed in more detail in Appendix \ref{app:cruv}.

By contrast, we confirmed that the impact of the initial \ce{H2} OPR on the water D/H ratio is negligible, likely because the \ce{H2} OPR becomes low ($<10^{-3}$) before the core stage, where most of the deuterium enrichment of water occurs.
The decrease in $\alpha$ from 0.95 to 0.90 leads to enhanced water D/H ratios because larger fraction of water ice forms during the core stage in the lower $\alpha$ models, while the overall trend is similar between the models with $\alpha= 0.95$ and $\alpha = 0.9$ (see Figure \ref{fig:HDO_H2O_h2opr1e-2_alpha010}).

\section{Discussion} \label{sec:dicuss}
\subsection{Conditions for reproducing the high water D/H ratio in 3I/ATLAS}
In Figs.\ref{fig:HDO_H2O__h2opr1e-2_alpha005} and \ref{fig:HDO_H2O_h2opr1e-2_alpha010}, red outlines indicate parameter sets where the water D/H ratio is consistent with 3I/ATLAS (allowing 30 \% deviation from 0.01), while blue outlines indicate parameter sets consistent with the ranges observed in nearby low-mass protostars (between $5\times10^{-4}$ and $3\times10^{-3}$).
Allowing 30 \% deviation was arbitrarily chosen but partly takes into account the uncertainty in the elemental D/H ratio, which can enhance the water D/H ratio, and the potential reprocessing during the disk formation and in the disk, which can reduce the water D/H ratio \citep[e.g.,][]{yang13,furuya17}.
With parameters representative of the local ISM ($Z=1$, $\chi = 1$, and $\zeta = 10^{-17}$ s$^{-1}$), our model reproduces the water D/H ratios observed in nearby low-mass protostars, supporting the validity of our modeling approach. 
This agreement holds over broad ranges of $\chi$ and $\zeta$, particularly when $n_{\rm cl} = 10^3$ cm$^{-3}$.

In our models, the observed water D/H ratio of of 3I/ATLAS, $\sim$0.01, is most readily reproduced at subsolar metallicities, $Z\lesssim0.5$, and relatively high cloud densities, $n_{\rm cl}\sim10^4\ {\rm cm}^{-3}$.
At $Z=0.5$, this ratio is reproduced in a restricted region of parameter space, whereas at $Z=0.2$ the allowed parameter space is broader.
For $Z=0.2$ and $n_{\rm cl}=10^4$ cm$^{-3}$, $\zeta=10^{-16}$ s$^{-1}$, above the standard dense ISM value (around a few $\times 10^{-17}$ s$^{-1}$; \citealt{padovani18,obolentseva24}), are slightly favored.
This is consistent with the enhanced-ionization scenario suggested by \citet{cordiner26} for the low-metallicity environment of 3I/ATLAS.
However, the ranges of $\zeta$ and $\chi$ that reproduce the observations shift with other parameters, such as $\alpha$, making them difficult to constrain solely from the water D/H ratio (see Fig. \ref{fig:HDO_H2O_h2opr1e-2_alpha010}).
Ionization rates of $\zeta\gtrsim10^{-15}$ s$^{-1}$ are disfavored at all metallicities, because cosmic-ray heating suppresses deuterium fractionation, as discussed above.

As an exception, even at $Z=1$, a water D/H ratio above 0.01 is obtained for a specific parameter set, $n_{\rm cl}=10^4\ {\rm cm}^{-3}$, $\chi=10$, and $\zeta=10^{-16}\ {\rm s}^{-1}$ (see Appendix \ref{app:cruv}).

Overall, the conditions that favor the high water D/H ratio of 3I/ATLAS in our models ($Z \lesssim 0.5$, $n_{\rm cl} \sim 10^4$ cm$^{-3}$, and 10$^{-17} \lesssim \zeta < 10^{-15}$ s$^{-1}$) differ from those that reproduce the water D/H ratios observed in nearby protostars ($Z = 1$, $n_{\rm cl} \sim 10^3$-10$^4$ cm$^{-3}$, and  $\zeta < 10^{-15}$ s$^{-1}$).
Although $n_{\rm cl}$ and $Z$ are treated as independent parameters in our models, they might be correlated in actual star-forming regions.
Because lower $Z$ requires a larger gas column density to provide sufficient shielding for water ice formation, lower $Z$ environments may preferentially reach the relevant shielding conditions in denser regions.

\begin{figure*}[ht!]
\includegraphics[width=\linewidth]{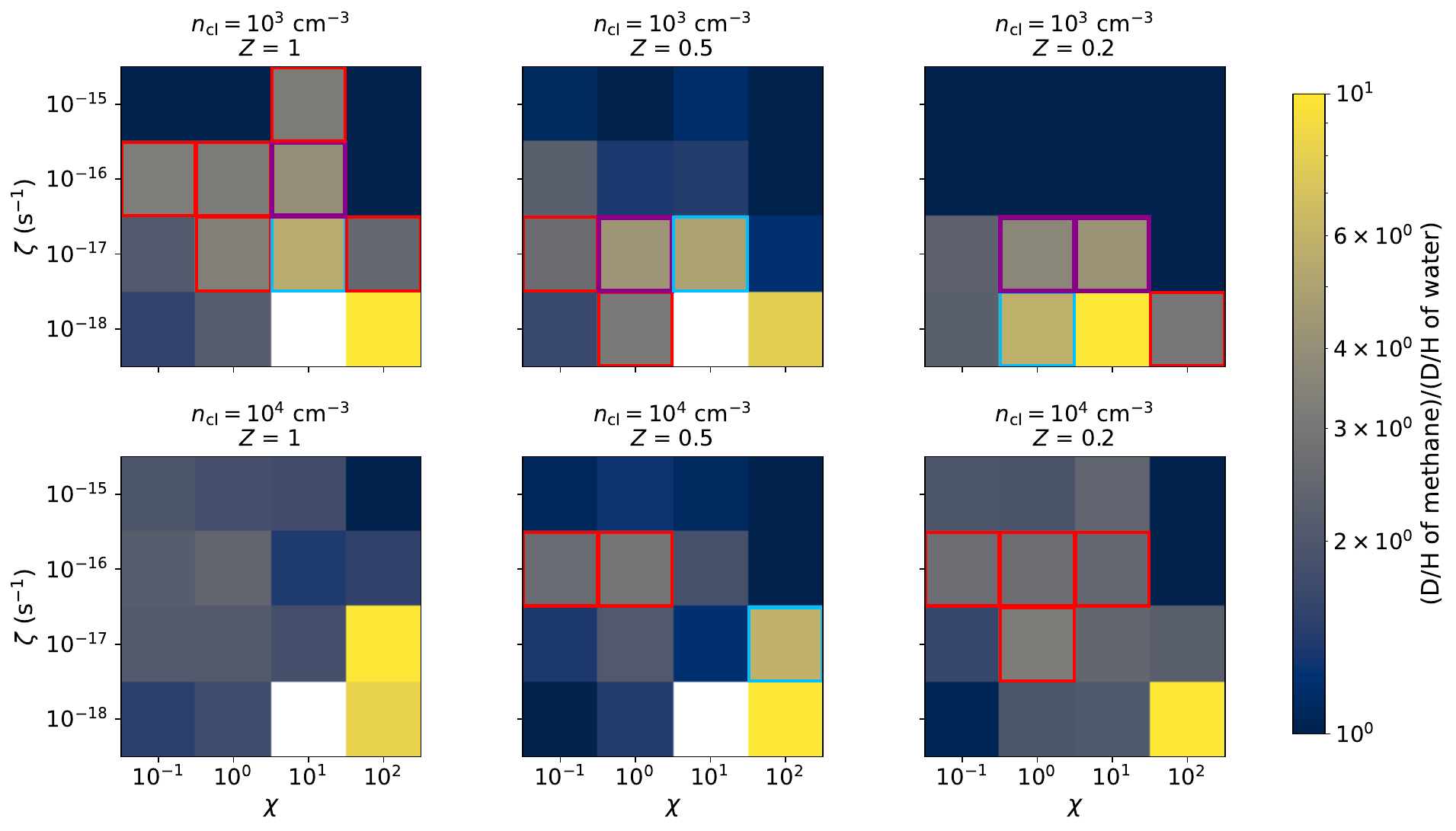}
\caption{Similar to Fig. \ref{fig:HDO_H2O__h2opr1e-2_alpha005}, but for the methane D/H ratio normalized by the water D/H ratio.
Red and blue outlines highlight parameter sets for which the normalized ratio is close to that of 3I/ATLAS and 67P/C-G, respectively (allowing a 30 \% deviation from 3.5 and 4.8). Magenta outlines indicate parameter sets whose normalized ratios are consistent with both 3I/ATLAS and 67P/C-G within the adopted tolerance.
}
\label{fig:CH3D_CH4_vs_HDO_H2O_2e4_5}
\end{figure*}

\subsection{D/H ratio of methane}
In addition to the high water D/H ratio, \citet{roth26} reported a very high methane D/H ratio of $\sim$3 \% in 3I/ATLAS, compared with $\sim$0.2 \% in the Solar System comet 67P/Churyumov-Gerasimenko (hereafter, 67P/C-G; \citealt{muller22}).
The methane D/H ratio has not been measured in other Solar System comets or in star-forming regions.
As pointed out by \citet{roth26}; however, the relative level of deuteration between methane and water is more comparable between the two comets, $3.4 \pm 0.4$ in 3I/ATLAS and $4.8 \pm 0.7$ or $9.3 \pm 1.7$ in 67P/C-G, depending on the adopted water D/H ratio \citep{muller22,mandt24}.

Figure \ref{fig:CH3D_CH4_vs_HDO_H2O_2e4_5} shows the ratio between the D/H ratios of methane and water (methane-to-water D/H ratio) predicted by our subgrid of models.
Regardless of $Z$, methane ice and its deuterated isotopologues form through subsequent hydrogen and deuterium addition reactions to atomic C on grain surfaces. 
Unlike the absolute water D/H ratio, the methane-to-water D/H ratio is relatively insensitive to metallicity, because the metallicity-dependent effects on the water and methane D/H ratios cancel out, at least partly.
This behavior may explain why 3I/ATLAS and 67P/C-G exhibit methane-to-water D/H ratios of the same order, despite their different absolute D/H ratios and likely different formation environments.
The specific parameter set, $n_{\rm cl}=10^4\ {\rm cm}^{-3}$, $\chi=10$, and $\zeta=10^{-16}\ {\rm s}^{-1}$, which reproduces the water D/H ratio above 0.01 even at $Z=1$, underpredicts the methane-to-water D/H ratio compared to the observed value.

Overall, our models suggest that the extreme water D/H ratio of 3I/ATLAS is most readily explained by formation in a low-metallicity, relatively high-density environment, assuming that the observed water was largely inherited from the parent molecular cloud and core. 
Its water D/H ratio therefore provides a constraint complementary to that from the high \ce{^12C}/\ce{^13C} ratio.
Measurements of water D/H ratios in low-metallicity star-forming regions will be crucial for testing this interpretation.

\acknowledgments
We would like to thank the anonymous referee for their prompt and constructive comments.
This work is, in part, supported by JSPS KAKENHI Grant numbers JP25K07364 and JP26H02071. 
MAC and NXR were supported by NASA’s Planetary Science Division Internal Scientist Funding Program through the Fundamental Laboratory Research work package (FLaRe). 
C.O.C. is supporteded by Schmidt Sciences. C.O.C. gratefully acknowledges support from the NASA CSSFP (grant No. 80NSSC26K0380).
MND is funded by the European Union. Views and opinions expressed are however those of the author(s) only and do not necessarily reflect those of the European Union or the European Research Council Executive Agency. Neither the European Union nor the granting authority can be held responsible for them. MND's work is supported by ERC grant PSII (DOI: 10.3030/101230593).
Numerical computations were in part carried out on a PC cluster at the Center for Computational Astrophysics, National Astronomical Observatory of Japan.

\software{Matplotlib \citep{matplotlib}}

\clearpage

\renewcommand{\thefigure}{A\arabic{figure}}
\setcounter{figure}{0}
\renewcommand{\thetable}{A\arabic{table}}
\setcounter{table}{0}

\begin{appendix}
\section{The steady-state abundance of atomic H at various  metallicities} \label{app:atomicH}
When UV photodissociation of \ce{H2} is negligible due to self-shielding, the steady-state abundance of atomic H in the gas phase is mostly determined by the balance between the cosmic-ray ionization of \ce{H2}, followed by the ion-neutral reaction between \ce{H2+} and \ce{H2}, and adsorption of atomic H onto dust grains, followed by their consumption via grain surface two-body reactions \citep[e.g., \ce{H2} formation via recombination of two H atoms;][]{goldsmith05}:
\begin{align}
\zeta n(\ce{H2}) &= fS_{\rm H}\sigma_{\rm gr}v_{\rm th, \, H}n_{\rm gr}n(\ce{H}),  \label{eq:eq_ab_H}
\end{align}
where $n(\ce{H2})$, $n(\ce{H})$, $n_{\rm gr}$ are the number density of \ce{H2}, atomic H and dust grains, respectively, per unit gas volume.
It is assumed that $n_{\rm gr}$ is proportional to $Z$ as in our numerical models.
$S_{\rm H}$ is the sticking probability of atomic H onto grains \citep{hollenbach79}, $f$ is the fraction of adsorbed atomic H on grains that is consumed by grain surface reactions, $\sigma_{\rm gr}$ is the cross section of a single dust grain, and $v_{\rm th, \, H}$ is the thermal velocity of atomic H.
Then the atomic H abundance is estimated as
\begin{align}
x_{\rm CR}(\ce{H}) &\approx 2 \times 10^{-5} \left( \frac{\zeta/n_{\ce{H}}}{5 \times 10^{-22}\,\,{\rm cm^3 \,\, s^{-1}}} \right)
\left( \frac{Z}{1} \right)^{-1}
\left( \frac{f}{1} \right)^{-1}
\left( \frac{S_{\rm H}}{0.8} \right)^{-1}, \label{eq:ab_H}
\end{align}
where $n_{\ce{H}}$ is the number density of hydrogen nuclei ($2n(\ce{H2}) + n(\ce{H})$).
However, in our models, the gas-phase abundance of atomic H is higher than that predicted by Eq. \ref{eq:ab_H} at the end of cloud stage (see upper left panel of Fig. \ref{fig:time_evol}).
In addition, although Eq. \ref{eq:ab_H} predicts a higher atomic H abundance at lower $Z$, the numerical results do not show this trend.
In the core stage, by contrast, the atomic H abundance largely follows the expectation from Eq. \ref{eq:ab_H} (bottom left panel of Fig. \ref{fig:time_evol}).

As discussed in Section \ref{sec:result}, the release of atomic H into the gas phase upon water ice photodissociation is an important source of atomic H in the cloud stage, where the external UV radiation is not fully shielded.
We consider a situation in which the \ce{H2} photodissociation is negligible due to self-shielding, while \ce{H2O} ice photodissociation and photodesorption are not negligible.
In this case, the balance equation is modified to
\begin{align}
\zeta n(\ce{H2})  + b_r R_{\rm phdiss} &= fS_{\rm H}\sigma_{\rm gr}v_{\rm th, \, H}n_{\rm gr}n(\ce{H}),
\end{align}
where $R_{\rm phdiss}$ is the total photodissociation rate of water ice and $b_r$ is the branching ratio for the release of atomic H into the gas phase as one of outcomes of water ice photodissociation.
As water ice photodesorption is also one possible outcomes of water ice photodissociation, and its branching ratio is $\sim$2 \% \citep{arasa15}, $R_{\rm phdiss}$ can be rewritten as $\sim$50$R_{\rm phdes}$, where $R_{\rm phdes}$ is the photodesorption rate of water ice.
By the definition of $A_{V, crit}$, $R_{\rm phdes}$ at $A_V = A_{V, crit}$ is equal to the adsorption rate of atomic O on dust grains.
Taken together, the atomic H abundance at $A_V = A_{V, crit}$ is estimated as 
\begin{align}
x(\ce{H}) &\approx x_{\rm CR}(\ce{H}) + 10^{-
4}\left( \frac{1-\alpha}{0.05} \right) \left( \frac{x_{\rm O}}{1.8\times10^{-4}} \right) \left( \frac{Z}{1} \right) \left( \frac{f}{1} \right)^{-1} 
\left( \frac{S_{\rm H}}{0.8} \right)^{-1}, \label{eq:ab_H_mod}
\end{align}
where the sticking probability of atomic O is set to be unity.
The second term on the right-hand side of Eq. \ref{eq:ab_H_mod} is proportional to $Z$, in contrast to the first term.
This is because the adsorption rate of atomic O scales with $Z^2$.
Figure \ref{eq:ab_H_mod} compares the atomic H abundance at $A_{V}=A_{V, crit}$ predicted by Eq. \ref{eq:ab_H_mod} with that predicted by Eq. \ref{eq:ab_H}.
Including the water ice photodissociation increases the atomic H abundance.
It also reverses the metallicity dependence of the atomic H abundance when $\zeta/n_{\rm H} \lesssim 5\times10^{-22}$ cm$^3$ s$^{-1}$, in agreement with our numerical results.

Similarly, atomic D can be released into the gas phase upon the photodissociation of HDO and \ce{D2O} ices \citep{arasa15}.
However, the impact of the atomic D release into the gas phase is smaller compared to the corresponding atomic H release, because the D/H ratio of water ice surface is much lower than the gas-phase atomic D/H ratio in the cloud stage.
Therefore, the net effect is to decrease the gas-phase atomic D/H ratio, and is more significant at higher $Z$.
Note that even when $Z=0.2$, where the impact of the atomic H release is small, the atomic D/H ratio is lower than the D/H ratio of \ce{H3+} by a factor of $\sim$3 in the cloud stage (panel (d) in Fig. \ref{fig:time_evol}).
This difference is attributed to two things. 
First, because of its larger mass, atomic D has a lower thermal velocity and thus a lower adsorption rate coefficient onto grains than atomic H. Second, atomic D is produced mainly by electron recombination of deuterated ions, whereas atomic H has additional gas-phase formation pathways, such as cosmic-ray-induced dissociation of \ce{H2}.

\begin{figure}[h]
\begin{center}
\includegraphics[width=0.5\linewidth]{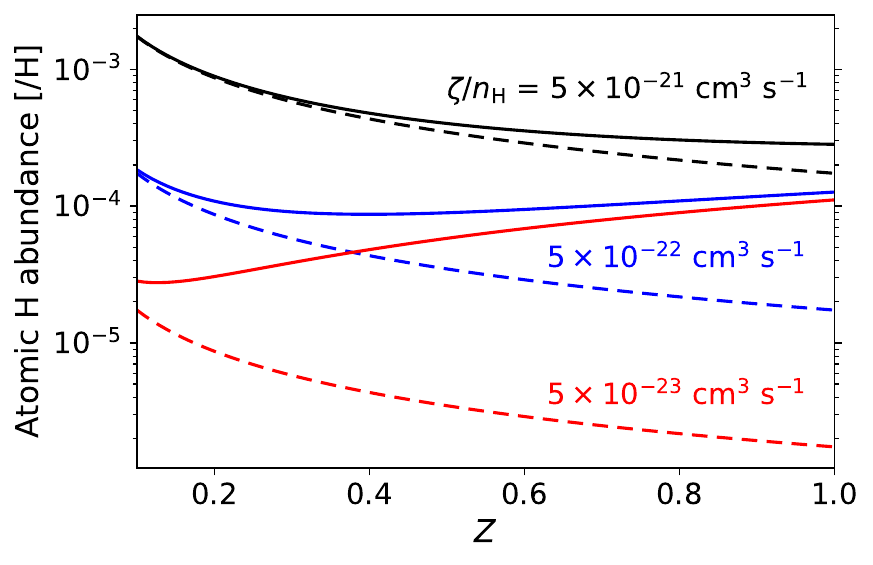}
\end{center}
\caption{The predicted gas-phase atomic H abundance with respect to hydrogen nuclei at $A_{V}=A_{V, crit}$ as functions of $Z$ and the $\zeta/n_{\rm H}$. 
Solid lines and dashed lines represent with (Eq. \ref{eq:ab_H_mod}) and without (Eq. \ref{eq:ab_H}) the atomic H release to the gas phase upon water ice photodissociation.
The parameter $\alpha$ is set to 0.95.}
\label{fig:hab_analytical}
\end{figure}

\section{Parameter dependence of the water D/H ratio} \label{app:cruv}
\subsection{Elemental D/H ratio} \label{app:elemental_dh}
The primordial elemental D/H ratio established by big bang nucleosynthesis is $2.5\times10^{-5}$ \citep{cooke18}.
Because deuterium is destroyed through stellar astration, the elemental D/H ratio in the Galaxy generally decreases over cosmic time, although inflow of unprocessed and deuterium-rich gas which has experienced little astration can modify this trend \citep{romano06}.
In our models, however, elemental D/H ratio is fixed to $1.5\times10^{-5}$ \citep{linsky03}, independent of metallicity.
To check the impact of the assumed elemental D/H ratio, panels (d) and (h) of Figure \ref{fig:time_evol_ncl} show the results obtained with a higher elemental D/H ratio of $2.5\times10^{-5}$ and $Z=0.2$.
These panels can be compared with panels (d) and (h) of Fig. \ref{fig:time_evol}; for ease of comparison, the latter are reproduced as panels (c) and (g) of Fig. \ref{fig:time_evol_ncl}.
We confirmed that the water D/H ratio almost linearly scales with the assumed elemental D/H ratio.
Then our models with subsolar metallicities can underestimate the water D/H ratio by a factor of $\sim$1.7 at the maximum.
This uncertainty does not affect our primary conclusion that the subsolar metallicity is favorable for reproducing the high water D/H ratio measured in 3I/ATLAS.

\subsection{Cloud density}
\begin{figure*}[ht!]
\begin{center}
\includegraphics[width=\linewidth]{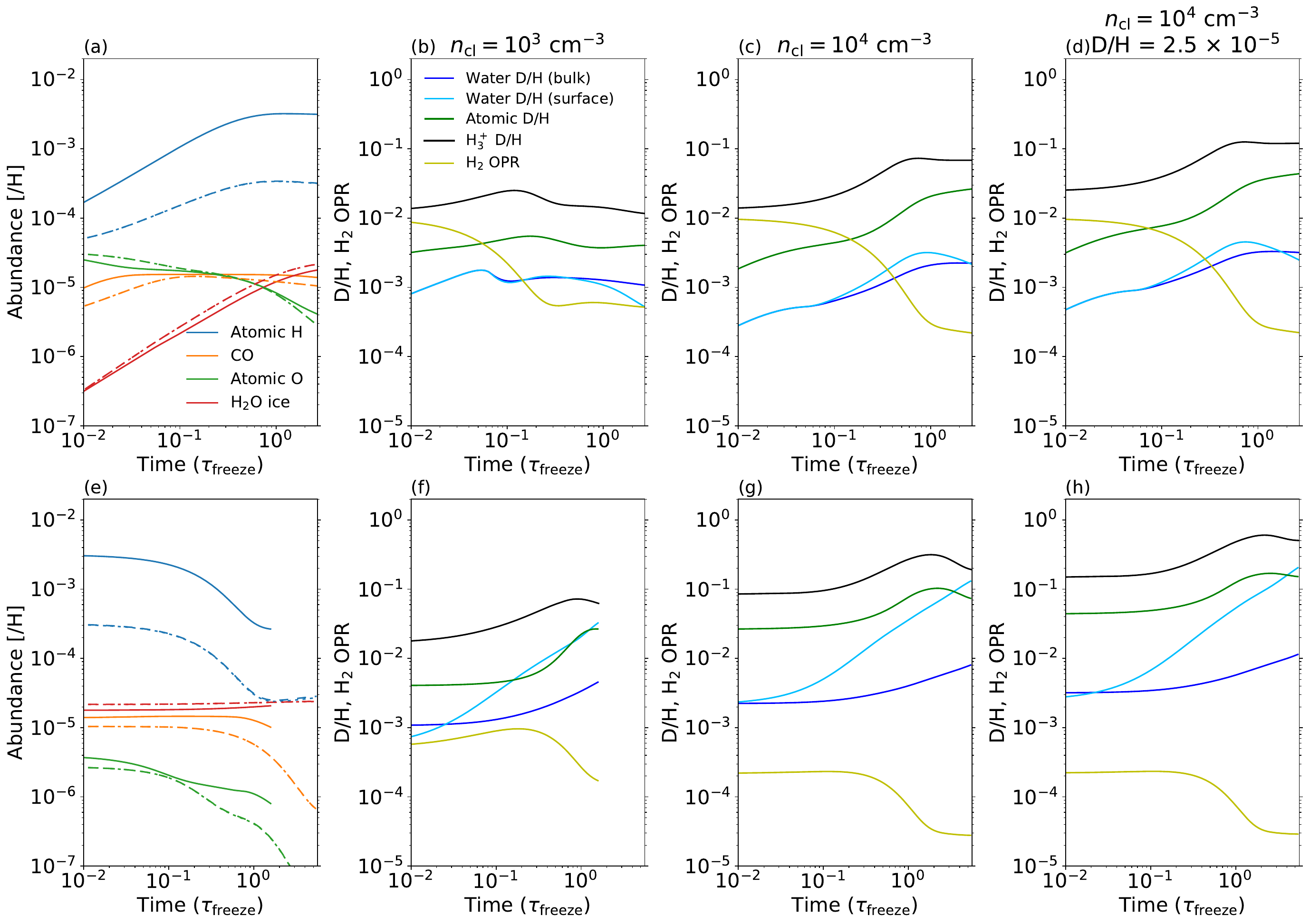}
\caption{Similar to Fig. \ref{fig:time_evol}, but for different $n_{\rm cl}$ values rather than $Z$. In the left panels, solid and dashed lines represent the models with $n_{\rm cl} = 10^{3}$ cm$^{-3}$ and 10$^{4}$ cm$^{-3}$, respectively.
In all the models, $Z$ is set to 0.2.}
\label{fig:time_evol_ncl}
\end{center}
\end{figure*}

Panels (b) and (f) of Fig. \ref{fig:time_evol_ncl}  show the temporal evolution of the water D/H ratio during the cloud and core stages, respectively, in the models with $n_{\rm cl}$ = 10$^3$ cm$^{-3}$ and with $Z = 0.2$. The other parameters are fixed at the fiducial values listed in Table \ref{table:parameter}.
Compared to the model with $n_{\rm cl}$ = 10$^4$ cm$^{-3}$ (panels (c) and (g)), the abundance of atomic H is higher (see also Fig. \ref{fig:hab_analytical}), resulting in a lower atomic D/H ratio in the gas phase, because the maximum abundance of atomic D is limited to $1.5 \times 10^{-5}$ (i.e., the elemental abundance of D).
As a result, the water D/H ratio at the end of core stage is lower in the lower cloud density model.

\subsection{Cosmic-ray ionization rate}
\begin{figure*}[ht!]
\includegraphics[width=\linewidth]{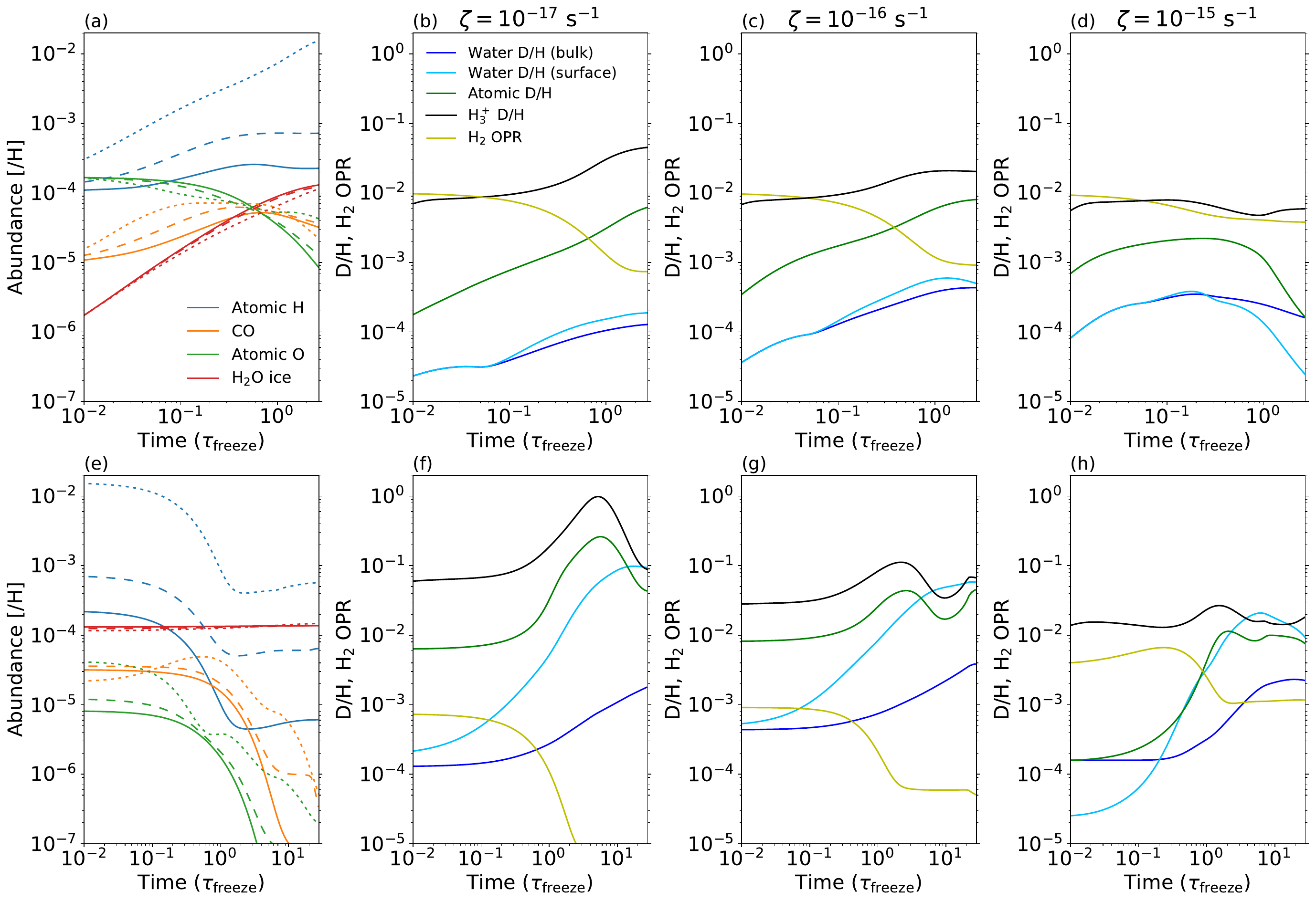}
\caption{Similar to Fig. \ref{fig:time_evol}, but for different $\zeta$ values rather than $Z$. In the left panels, solid, dashed, and dotted lines represent the models with $\zeta = 10^{-17}$, 10$^{-16}$, and 10$^{-15}$ s$^{-1}$, respectively.
In all the models, $Z$ is set to unity.}
\label{fig:time_evol_cr}
\end{figure*}

Figure \ref{fig:time_evol_cr} is similar to Figure \ref{fig:time_evol}, but shows models with different values of $\zeta$.
In all models shown in the figure, $Z = 1$.
The temporal evolution of the water ice abundance is similar regardless of $\zeta$.
The gas temperature in the cloud stage is $\sim$10, $\sim$13, and $\sim$22 K for $\zeta=10^{-17}$, $10^{-16}$, and $10^{-15}$ s$^{-1}$, respectively, making deuterium fractionation through Reaction \ref{react1} less efficient at higher $\zeta$.
This effect is particularly important when $\zeta = 10^{-15}$ s$^{-1}$. 
The abundance of atomic H increases with increasing $\zeta$ (see also Fig. \ref{fig:hab_analytical}), resulting in a lower atomic D/H ratio in the gas phase.
On the other hand, the enhanced $\zeta$ promotes the reaction CO + \ce{He+} $\rightarrow$ \ce{C+} + O, providing the additional source of oxygen for the formation of deuterated water ice during the core stage (i.e., effectively reduces the parameter $\alpha$).
As a result of these competing effects, the water D/H ratio at the end of the core stage is highest in the model with $\zeta = 10^{-16}$ s$^{-1}$ among the three cases considered.

\subsection{External UV radiation}
\begin{figure*}[ht!]
\includegraphics[width=\linewidth]{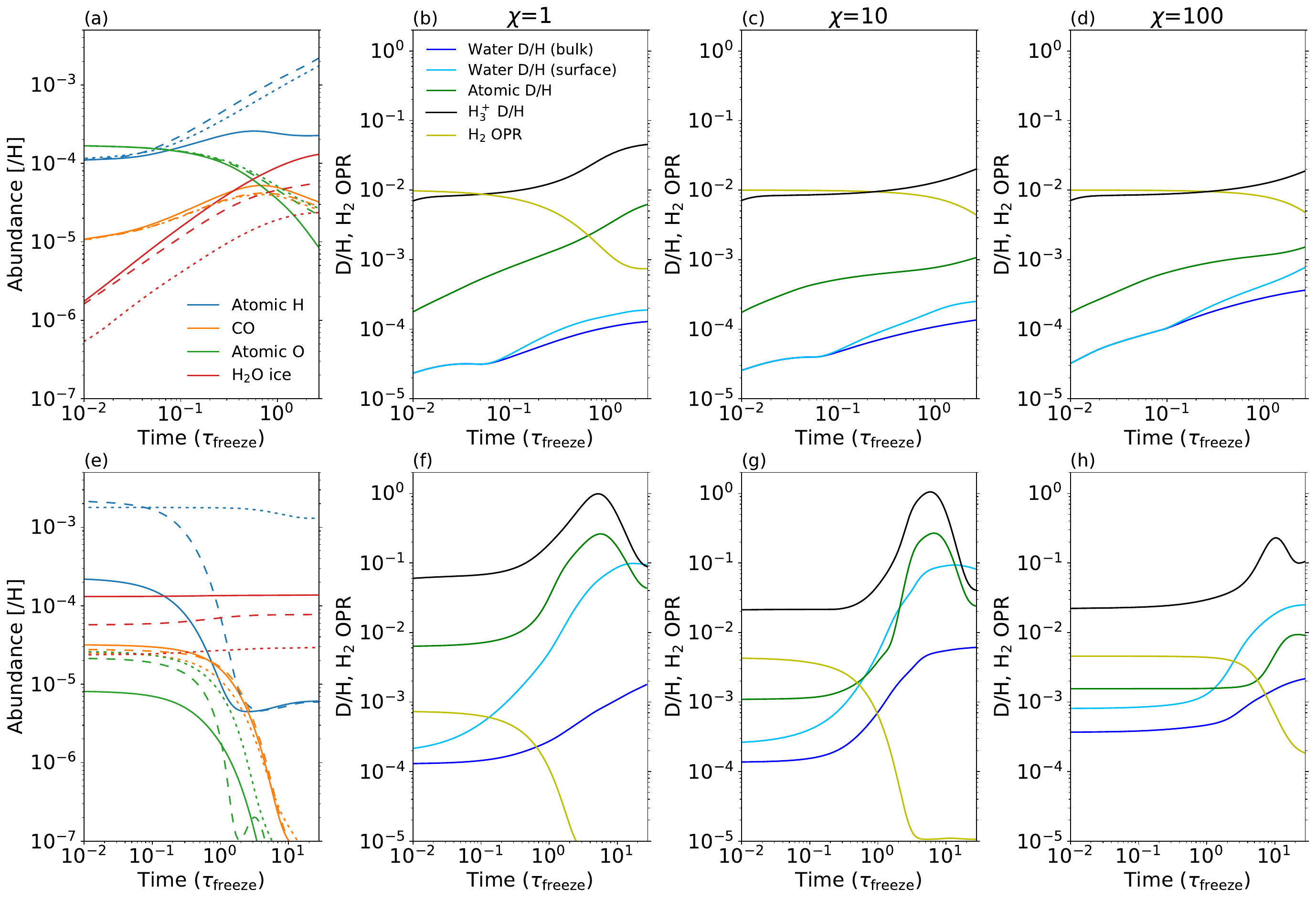}
\caption{Similar to Fig. \ref{fig:time_evol}, but for different $\chi$ values rather than $Z$.
In the left panels, solid, dashed, and dotted lines represent the models with $\chi = 1$, 10, and 100, respectively. In all the models, $Z$ is set to unity.}
\label{fig:time_evol_uv}
\end{figure*}

Figure \ref{fig:time_evol_uv} is similar to Figure \ref{fig:time_evol_ncl}, but shows models with different values of $\chi$. 
In all models shown in the figure, $Z = 1$.
As noted in Sec. \ref{sec:modelsetup}, the UV field strength relevant to photochemistry and gas heating does not strongly depend on $\chi$ during the cloud stage in our models, while the dust temperature increases with increasing $\chi$.
The dust temperature in the cloud stage is 12, 16, and 21 K for $\chi=1$, $10$, and $100$, respectively, while the gas temperature is $\sim$10 K regardless of $\chi$.
Due to the higher dust temperature, the water ice abundance decreases with increasing $\chi$, while the \ce{CO2} ice abundance increases (not shown in the figure).
The enhanced mobility of CO on grain surfaces leads to more efficient \ce{CO2} ice formation through reactions between CO and OH, which competes with water ice formation.
The higher dust temperature also enhances thermal desorption of atomic H from grain surfaces, leading to the reduced fraction of adsorbed atomic H consumed by grain surface reactions (i.e., the parameter $f$ in Eq. \ref{eq:ab_H} becomes smaller).
As a result, the gas-phase abundance of atomic H increases with increasing $\chi$, leading to a lower atomic D/H ratio, as in the high $\zeta$ models.
On the other hand, at higher dust temperatures, the grain surface reaction OH + H $\rightarrow$ \ce{H2O} becomes more important for water ice formation than OH + \ce{H2} $\rightarrow$ \ce{H2O} + H.
This change in the formation pathway occurs because the binding energy of atomic H and \ce{H2} are assumed to be the same, and because the reaction OH + \ce{H2} has an activation energy barrier.
The increased relative contribution of the OH + H pathway leads to a higher water D/H ratio \citep{furuya15}.
As a result of these competing effects, the water D/H ratio at the end of the core stage is the highest in the $\chi = 10$ model among the three cases considered.

\section{Additional figures for water D/H} \label{app:figures}
\begin{figure*}[ht!]
\includegraphics[width=\linewidth]{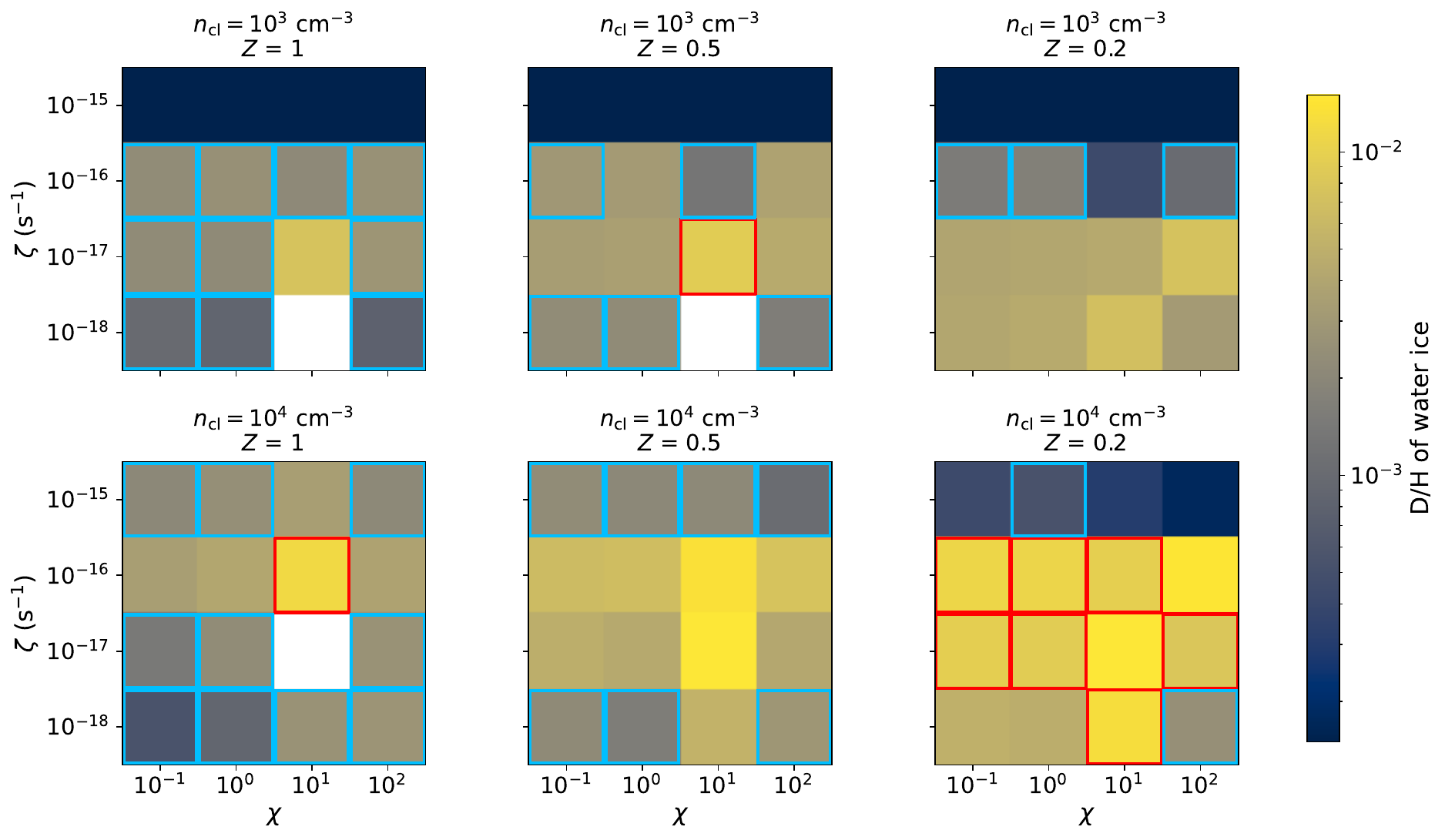}
\caption{Similar to Figure \ref{fig:HDO_H2O__h2opr1e-2_alpha005}, but the parameter $\alpha$ is set to 0.9.
}
\label{fig:HDO_H2O_h2opr1e-2_alpha010}
\end{figure*}

Figure \ref{fig:HDO_H2O_h2opr1e-2_alpha010} shows the water D/H ratio at the end of the core stage for models with fixed $\alpha = 0.9$ and an initial H$_2$ OPR of $10^{-2}$.
The overall trend is similar to Fig. \ref{fig:HDO_H2O__h2opr1e-2_alpha005}, where $\alpha = 0.95$ and an initial H$_2$ OPR of $10^{-2}$.

\begin{figure*}[ht!]
\includegraphics[width=\linewidth]{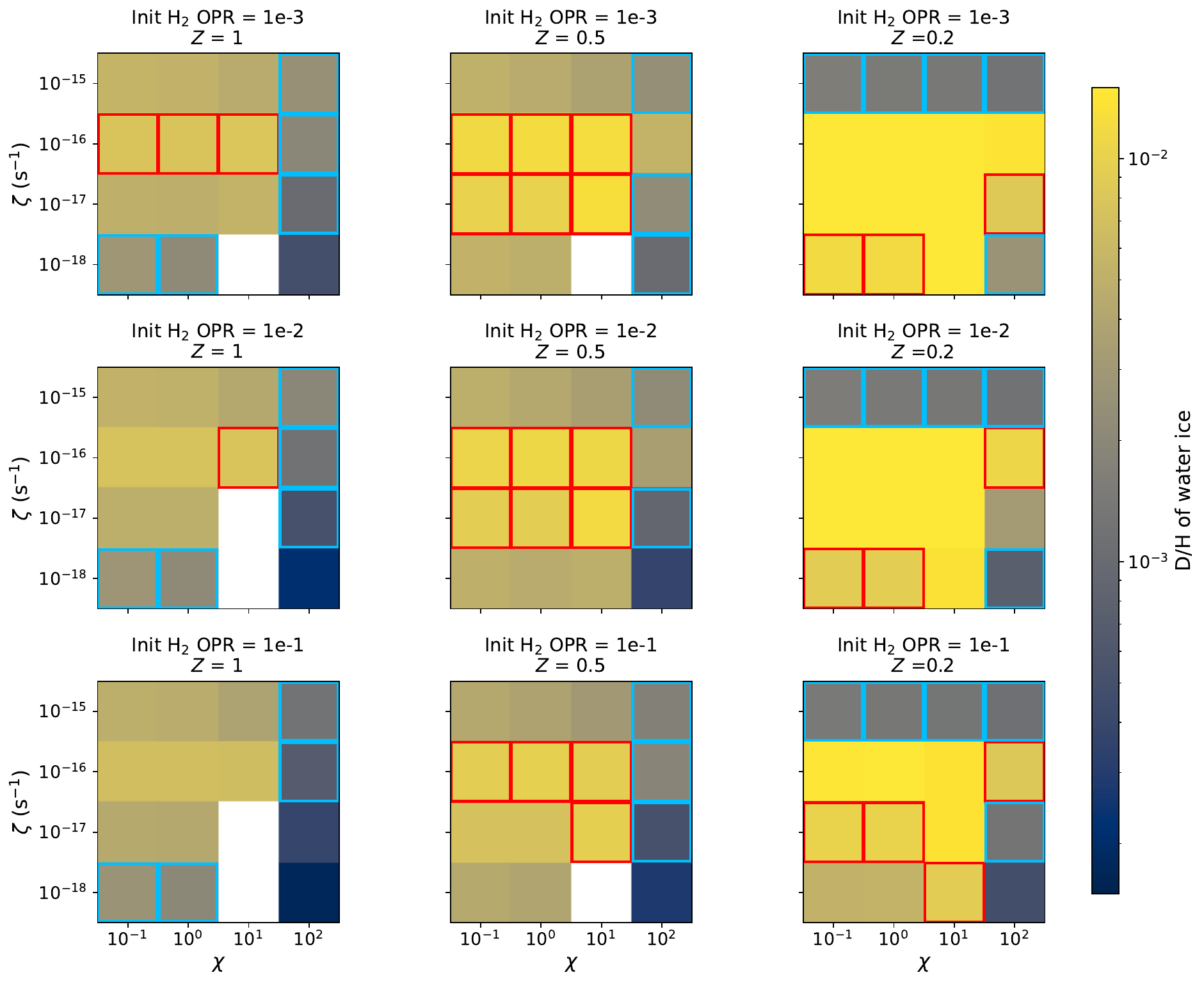}
\caption{Similar to Fig. \ref{fig:HDO_H2O__h2opr1e-2_alpha005}, but for the models neglecting the cloud phase (i.e., $\alpha = 0$).
The top, middle, and bottom panels correspond to the initial \ce{H2} OPR of 10$^{-3}$, 10$^{-2}$, and 10$^{-1}$, respectively. From left to right, the metallicity decreases from $Z=1$ to 0.2. The core density is $10^5$ cm$^{-3}$ in all the models.
}
\label{fig:waterDH_2e5_coreonly}
\end{figure*}

Figure \ref{fig:waterDH_2e5_coreonly} shows the water D/H ratio predicted by additional models in which the cloud phase is neglected and the chemical evolution starts from the core phase (i.e., $\alpha=0$), where the external UV radiation is negligible.
These models can be compared with previous astrochemical models that did not consider the cloud phase \citep[e.g.,][]{aikawa12,lee15}.
However, such models may be less realistic, given that \ce{H2O} ice is already abundant in relatively low-$A_V$ regions, with line-of-sight extinctions of $A_V\sim3$ mag in nearby molecular clouds \citep[e.g.,][]{boogert15}.
Even at $Z=1$, the water D/H ratio can reach $\sim0.01$, consistent with previous models.
In our model, this occurs particularly when the initial \ce{H2} OPR is $10^{-3}$ and $\zeta=10^{-16}$ s$^{-1}$. 
Even in this more simplified setup, the water D/H ratio is higher at lower $Z$.
Since the effect of water ice photodissociation is negligible, the higher water D/H ratio at lower $Z$ is simply attributed to the enhanced D/H ratio of \ce{H3+} at a given normalized time $t/\tau_{\rm freeze}$, 
because of a lower CO abundance, lower \ce{H2} OPR, and longer freeze-out timescale.

\section{D$_2$O} \label{app:d2o}
Although there is no constraint on \ce{D2O} toward 3I/ATLAS, it may be interesting to discuss how the \ce{D2O}/HDO ratio relative to HDO/\ce{H2O}, which has been proposed as a good probe of the prestellar inheritance of water \citep{furuya17}, depend on the metallicity. The \ce{D2O}/HDO ratio is found to be higher than the HDO/\ce{H2O} ratio in both low-mass protostellar sources in the solar neighborhood and in the Solar System comet 67P/C-G \citep{coutens14,altwegg17,jensen21,leemker25}. To establish the higher \ce{D2O}/HDO than HDO/\ce{H2O}, the large gradient of water deuteration is required between the cloud phase and the core phase at least by a factor of $>$10 \citep{furuya16}. 

\begin{figure*}[ht!]
\includegraphics[width=\linewidth]{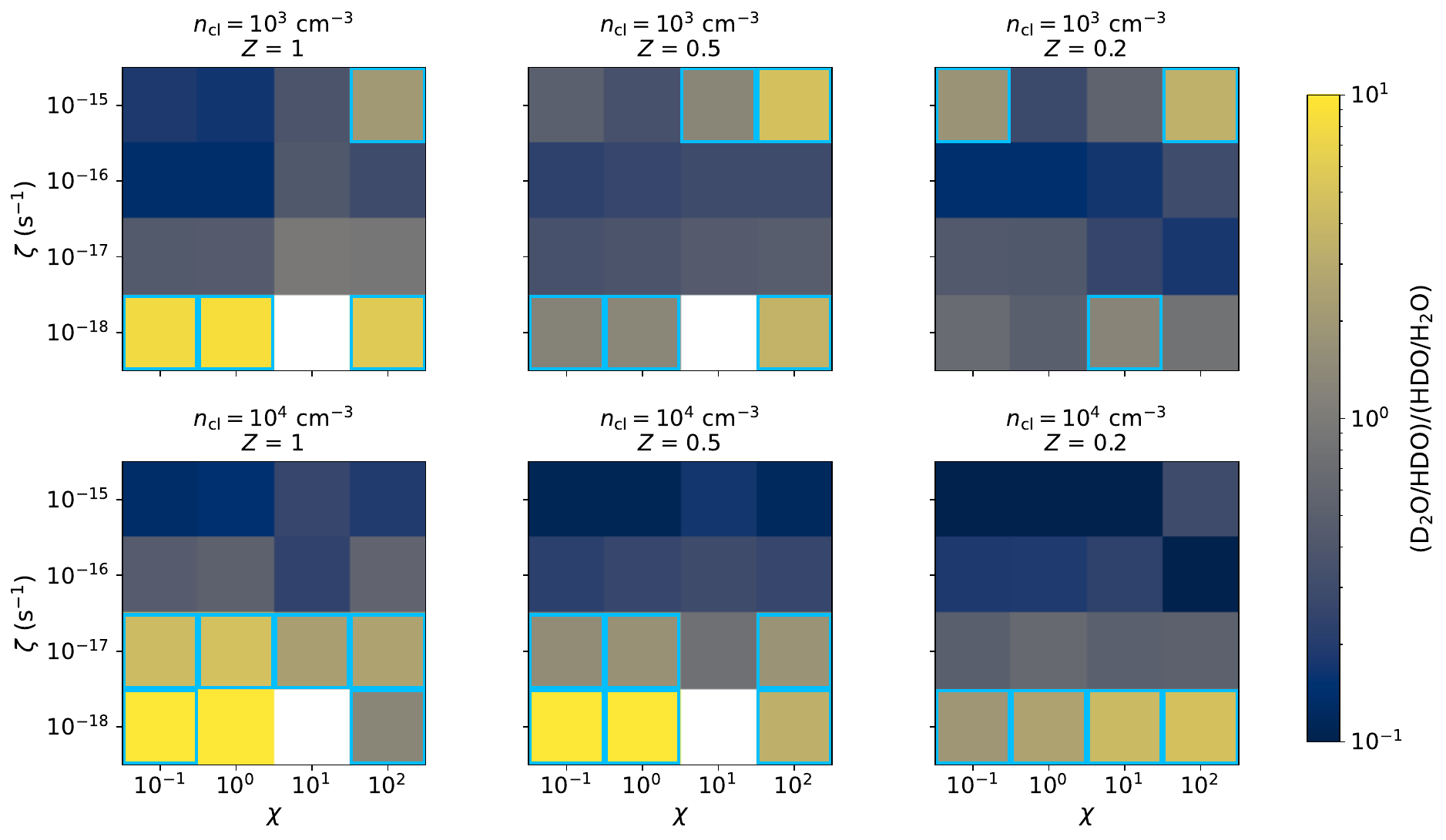}
\caption{Similar to Fig. \ref{fig:HDO_H2O__h2opr1e-2_alpha005}, but for the \ce{D2O}/HDO ratio with respect to the HDO/\ce{H2O} ratio. Blue outlines indicate parameter sets with the ratio is higher than unity as observed in nearby low-mass protostars and comet 67P/C-G.}
\label{fig:d2o}
\end{figure*}

At lower $Z$, the water deuteration in the cloud phase is more efficient and thus the deuteration gradient between the cloud stage and the core stage is smaller, leading to lower \ce{D2O}/HDO ratio relative to HDO/\ce{H2O} as shown in Figure \ref{fig:d2o}.

\end{appendix}




\bibliography{ref}{}

@ARTICLE{cordiner26,
       author = {{Cordiner}, Martin and {Roth}, Nathan X. and {Micheli}, Marco and {Villanueva}, Geronimo and {Farnocchia}, Davide and {Charnley}, Steven and {Biver}, Nicolas and {Bockel{\'e}e-Morvan}, Dominique and {Bodewits}, Dennis and {Chandler}, Colin Orion and {Crovisier}, Jacques and {Drozdovskaya}, Maria N. and {Furuya}, Kenji and {Kelley}, Michael S.~P. and {Milam}, Stefanie and {Noonan}, John W. and {Opitom}, Cyrielle and {Schwamb}, Megan E. and {Thomas}, Cristina A.},
        title = "{Isotopic evidence for a cold and distant origin of 3I/ATLAS}",
      journal = {\nat},
         year = 2026,
        month = jul,
       volume = {655},
       number = {8124},
        pages = {870-874},
          doi = {10.1038/s41586-026-10771-6},
archivePrefix = {arXiv},
       eprint = {2603.06911},
 primaryClass = {astro-ph.EP},
       adsurl = {https://ui.adsabs.harvard.edu/abs/2026Natur.655..870C}
}

@ARTICLE{bockelee-Morvan15,
       author = {{Bockel{\'e}e-Morvan}, Dominique and {Calmonte}, Ursina and {Charnley}, Steven and {Duprat}, Jean and {Engrand}, C{\'e}cile and {Gicquel}, Adeline and {H{\"a}ssig}, Myrtha and {Jehin}, Emmanu{\"e}l and {Kawakita}, Hideyo and {Marty}, Bernard and {Milam}, Stefanie and {Morse}, Andrew and {Rousselot}, Philippe and {Sheridan}, Simon and {Wirstr{\"o}m}, Eva},
        title = "{Cometary Isotopic Measurements}",
      journal = {\ssr},
         year = 2015,
        month = dec,
       volume = {197},
       number = {1-4},
        pages = {47-83},
          doi = {10.1007/s11214-015-0156-9},
       adsurl = {https://ui.adsabs.harvard.edu/abs/2015SSRv..197...47B}
}

@ARTICLE{romano06,
       author = {{Romano}, Donatella and {Tosi}, Monica and {Chiappini}, Cristina and {Matteucci}, Francesca},
        title = "{Deuterium astration in the local disc and beyond}",
      journal = {\mnras},
         year = 2006,
        month = jun,
       volume = {369},
       number = {1},
        pages = {295-304},
          doi = {10.1111/j.1365-2966.2006.10287.x},
archivePrefix = {arXiv},
       eprint = {astro-ph/0603190},
 primaryClass = {astro-ph},
       adsurl = {https://ui.adsabs.harvard.edu/abs/2006MNRAS.369..295R}
}

@ARTICLE{milam05,
       author = {{Milam}, S.~N. and {Savage}, C. and {Brewster}, M.~A. and {Ziurys}, L.~M. and {Wyckoff}, S.},
        title = "{The $^{12}$C/$^{13}$C Isotope Gradient Derived from Millimeter Transitions of CN: The Case for Galactic Chemical Evolution}",
      journal = {\apj},
         year = 2005,
        month = dec,
       volume = {634},
       number = {2},
        pages = {1126-1132},
          doi = {10.1086/497123},
       adsurl = {https://ui.adsabs.harvard.edu/abs/2005ApJ...634.1126M}
}

@ARTICLE{lyons18,
       author = {{Lyons}, James R. and {Gharib-Nezhad}, Ehsan and {Ayres}, Thomas R.},
        title = "{A light carbon isotope composition for the Sun}",
      journal = {Nature Communications},
         year = 2018,
        month = mar,
       volume = {9},
          eid = {908},
        pages = {908},
          doi = {10.1038/s41467-018-03093-3},
       adsurl = {https://ui.adsabs.harvard.edu/abs/2018NatCo...9..908L}
}

@ARTICLE{opitom26,
       author = {{Opitom}, C. and {Manfroid}, J. and {Hutsem{\'e}kers}, D. and {Jehin}, E. and {Knight}, M.~M. and {Aravind}, K. and {Ferellec}, L. and {Bodewits}, D. and {Guzm{\'a}n}, V.~V. and {Cordiner}, M. and {Dorsey}, R.~C. and {La Forgia}, F. and {Lippi}, M. and {Murphy}, B.~P. and {Snodgrass}, C. and {Bannister}, M.},
        title = "{High nitrogen and carbon isotopic ratios in the interstellar comet 3I/ATLAS}",
      journal = {Nature Astronomy},
         year = 2026,
        month = jul,
          doi = {10.1038/s41550-026-02921-7},
archivePrefix = {arXiv},
       eprint = {2603.07187},
 primaryClass = {astro-ph.EP},
       adsurl = {https://ui.adsabs.harvard.edu/abs/2026NatAs.tmp..145O}
}

@ARTICLE{leemker25,
       author = {{Leemker}, Margot and {Tobin}, John J. and {Facchini}, Stefano and {Curone}, Pietro and {Booth}, Alice S. and {Furuya}, Kenji and {van't Hoff}, Merel L.~R.},
        title = "{Pristine ices in a planet-forming disk revealed by heavy water}",
      journal = {Nature Astronomy},
         year = 2025,
        month = oct,
       volume = {9},
        pages = {1486-1494},
          doi = {10.1038/s41550-025-02663-y},
archivePrefix = {arXiv},
       eprint = {2510.19919},
 primaryClass = {astro-ph.EP},
       adsurl = {https://ui.adsabs.harvard.edu/abs/2025NatAs...9.1486L}
}

@ARTICLE{obolentseva24,
       author = {{Obolentseva}, M. and {Ivlev}, A.~V. and {Silsbee}, K. and {Neufeld}, D.~A. and {Caselli}, P. and {Edenhofer}, G. and {Indriolo}, N. and {Bisbas}, T.~G. and {Lomeli}, D.},
        title = "{Reevaluation of the Cosmic-Ray Ionization Rate in Diffuse Clouds}",
      journal = {\apj},
         year = 2024,
        month = oct,
       volume = {973},
       number = {2},
          eid = {142},
        pages = {142},
          doi = {10.3847/1538-4357/ad71ce},
archivePrefix = {arXiv},
       eprint = {2408.11511},
 primaryClass = {astro-ph.GA},
       adsurl = {https://ui.adsabs.harvard.edu/abs/2024ApJ...973..142O}
}

@ARTICLE{padovani18,
       author = {{Padovani}, Marco and {Ivlev}, Alexei V. and {Galli}, Daniele and {Caselli}, Paola},
        title = "{Cosmic-ray ionisation in circumstellar discs}",
      journal = {\aap},
         year = 2018,
        month = jun,
       volume = {614},
          eid = {A111},
        pages = {A111},
          doi = {10.1051/0004-6361/201732202},
archivePrefix = {arXiv},
       eprint = {1803.09348},
 primaryClass = {astro-ph.HE},
       adsurl = {https://ui.adsabs.harvard.edu/abs/2018A&A...614A.111P}
}

@ARTICLE{kalvans17,
       author = {{Kalv{\={a}}ns}, J. and {Shmeld}, I. and {Kalnin}, J.~R. and {Hocuk}, S.},
        title = "{Chemical fractionation of deuterium in the protosolar nebula}",
      journal = {\mnras},
         year = 2017,
        month = may,
       volume = {467},
       number = {2},
        pages = {1763-1775},
          doi = {10.1093/mnras/stx174},
archivePrefix = {arXiv},
       eprint = {1701.05856},
 primaryClass = {astro-ph.GA},
       adsurl = {https://ui.adsabs.harvard.edu/abs/2017MNRAS.467.1763K}
}

@ARTICLE{muller22,
       author = {{M{\"u}ller}, D.~R. and {Altwegg}, K. and {Berthelier}, J.~J. and {Combi}, M. and {De Keyser}, J. and {Fuselier}, S.~A. and {H{\"a}nni}, N. and {Pestoni}, B. and {Rubin}, M. and {Schroeder}, I.~R.~H.~G. and {Wampfler}, S.~F.},
        title = "{High D/H ratios in water and alkanes in comet 67P/Churyumov-Gerasimenko measured with Rosetta/ROSINA DFMS}",
      journal = {\aap},
         year = 2022,
        month = jun,
       volume = {662},
          eid = {A69},
        pages = {A69},
          doi = {10.1051/0004-6361/202142922},
archivePrefix = {arXiv},
       eprint = {2202.03521},
 primaryClass = {astro-ph.EP},
       adsurl = {https://ui.adsabs.harvard.edu/abs/2022A&A...662A..69M}
}

@ARTICLE{hollenbach09,
       author = {{Hollenbach}, David and {Kaufman}, Michael J. and {Bergin}, Edwin A. and {Melnick}, Gary J.},
        title = "{Water, O$_{2}$, and Ice in Molecular Clouds}",
      journal = {\apj},
         year = 2009,
        month = jan,
       volume = {690},
       number = {2},
        pages = {1497-1521},
          doi = {10.1088/0004-637X/690/2/1497},
archivePrefix = {arXiv},
       eprint = {0809.1642},
 primaryClass = {astro-ph},
       adsurl = {https://ui.adsabs.harvard.edu/abs/2009ApJ...690.1497H}
}

@ARTICLE{dalgarno84,
       author = {{Dalgarno}, A. and {Lepp}, S.},
        title = "{Deuterium fractionation mechanisms in interstellar clouds.}",
      journal = {\apjl},
         year = 1984,
        month = dec,
       volume = {287},
        pages = {L47-L50},
          doi = {10.1086/184395},
       adsurl = {https://ui.adsabs.harvard.edu/abs/1984ApJ...287L..47D}
}

@ARTICLE{jensen21,
       author = {{Jensen}, S.~S. and {J{\o}rgensen}, J.~K. and {Kristensen}, L.~E. and {Coutens}, A. and {van Dishoeck}, E.~F. and {Furuya}, K. and {Harsono}, D. and {Persson}, M.~V.},
        title = "{ALMA observations of doubly deuterated water: inheritance of water from the prestellar environment}",
      journal = {\aap},
         year = 2021,
        month = jun,
       volume = {650},
          eid = {A172},
        pages = {A172},
          doi = {10.1051/0004-6361/202140560},
archivePrefix = {arXiv},
       eprint = {2104.13411},
 primaryClass = {astro-ph.GA},
       adsurl = {https://ui.adsabs.harvard.edu/abs/2021A&A...650A.172J}
}

@ARTICLE{shimonishi21,
       author = {{Shimonishi}, Takashi and {Izumi}, Natsuko and {Furuya}, Kenji and {Yasui}, Chikako},
        title = "{The Detection of a Hot Molecular Core in the Extreme Outer Galaxy}",
      journal = {\apj},
         year = 2021,
        month = dec,
       volume = {922},
       number = {2},
          eid = {206},
        pages = {206},
          doi = {10.3847/1538-4357/ac289b},
archivePrefix = {arXiv},
       eprint = {2109.11123},
 primaryClass = {astro-ph.GA},
       adsurl = {https://ui.adsabs.harvard.edu/abs/2021ApJ...922..206S}
}

@ARTICLE{SalazarManzano26,
       author = {{Salazar Manzano}, Luis E. and {Paneque-Carre{\~n}o}, Teresa and {Cordiner}, Martin A. and {Bergin}, Edwin A. and {Lin}, Hsing Wen and {Lis}, Dariusz C. and {Gerdes}, David W. and {Bergner}, Jennifer B. and {Biver}, Nicolas and {Bockel{\'e}e-Morvan}, Dominique and {Bodewits}, Dennis and {Charnley}, Steven B. and {Crovisier}, Jacques and {Farnocchia}, Davide and {Guzm{\'a}n}, Viviana V. and {Milam}, Stefanie N. and {Noonan}, John W. and {Remijan}, Anthony J. and {Roth}, Nathan X. and {Tobin}, John J.},
        title = "{Water D/H in 3I/ATLAS as a probe of formation conditions in another planetary system}",
      journal = {Nature Astronomy},
         year = 2026,
        month = apr,
          doi = {10.1038/s41550-026-02850-5},
archivePrefix = {arXiv},
       eprint = {2603.07026},
 primaryClass = {astro-ph.EP},
       adsurl = {https://ui.adsabs.harvard.edu/abs/2026NatAs.tmp...89S}
}

@ARTICLE{altwegg17,
       author = {{Altwegg}, K. and {Balsiger}, H. and {Berthelier}, J.~J. and {Bieler}, A. and {Calmonte}, U. and {De Keyser}, J. and {Fiethe}, B. and {Fuselier}, S.~A. and {Gasc}, S. and {Gombosi}, T.~I. and {Owen}, T. and {Le Roy}, L. and {Rubin}, M. and {S{\'e}mon}, T. and {Tzou}, C.-Y.},
        title = "{D$_{2}$O and HDS in the coma of 67P/Churyumov-Gerasimenko}",
      journal = {Philosophical Transactions of the Royal Society of London Series A},
         year = 2017,
        month = may,
       volume = {375},
       number = {2097},
          eid = {20160253},
        pages = {20160253},
          doi = {10.1098/rsta.2016.0253},
       adsurl = {https://ui.adsabs.harvard.edu/abs/2017RSPTA.37560253A}
}

@ARTICLE{sewilo22,
       author = {{Sewi{\l}o}, Marta and {Karska}, Agata and {Kristensen}, Lars E. and {Charnley}, Steven B. and {Chen}, C.-H. Rosie and {Oliveira}, Joana M. and {Cordiner}, Martin and {Wiseman}, Jennifer and {S{\'a}nchez-Monge}, {\'A}lvaro and {van Loon}, Jacco Th. and {Indebetouw}, Remy and {Schilke}, Peter and {Garcia-Berrios}, Emmanuel},
        title = "{The Detection of Deuterated Water in the Large Magellanic Cloud with ALMA}",
      journal = {\apj},
         year = 2022,
        month = jul,
       volume = {933},
       number = {1},
          eid = {64},
        pages = {64},
          doi = {10.3847/1538-4357/ac6de1},
archivePrefix = {arXiv},
       eprint = {2205.04325},
 primaryClass = {astro-ph.GA},
       adsurl = {https://ui.adsabs.harvard.edu/abs/2022ApJ...933...64S}
}

@ARTICLE{roth26,
       author = {{Roth}, Nathan X. and {Cordiner}, Martin and {Milam}, Stefanie and {Villanueva}, Geronimo and {Charnley}, Steven and {Biver}, Nicolas and {Bockelee-Morvan}, Dominique and {Bodewits}, Dennis and {Crovisier}, Jacques and {Drozdovskaya}, Maria N. and {Farnocchia}, Davide and {Furuya}, Kenji and {Kelley}, Michael S.~P. and {Micheli}, Marco and {Noonan}, John W. and {Opitom}, Cyrielle and {Schwamb}, Megan E. and {Thomas}, Cristina A.},
        title = "{Isotopic Signature of Organic Molecules from Beyond the Solar System: An Enriched Methane D/H Ratio in the Interstellar Object 3I/ATLAS}",
      journal = {arXiv e-prints},
         year = 2026,
        month = mar,
          eid = {arXiv:2603.20445},
        pages = {arXiv:2603.20445},
          doi = {10.48550/arXiv.2603.20445},
archivePrefix = {arXiv},
       eprint = {2603.20445},
 primaryClass = {astro-ph.EP},
       adsurl = {https://ui.adsabs.harvard.edu/abs/2026arXiv260320445R}
}

@ARTICLE{furuya26,
       author = {{Furuya}, Kenji and {Sugimoto}, Toshiki and {Iwasaki}, Kazunari and {Tsuge}, Masashi and {Watanabe}, Naoki},
        title = "{H$_2$ Ortho-Para Spin Conversion on Inhomogeneous Grain Surfaces. II. impact of the rotational energy difference between adsorbed ortho-H$_2$ and para-H$_2$ and implication to deuterium fractionation chemistry}",
      journal = {arXiv e-prints},
         year = 2026,
        month = feb,
          eid = {arXiv:2602.13122},
        pages = {arXiv:2602.13122},
archivePrefix = {arXiv},
       eprint = {2602.13122},
 primaryClass = {astro-ph.GA},
       adsurl = {https://ui.adsabs.harvard.edu/abs/2026arXiv260213122F}
}

@ARTICLE{vandishoeck06,
       author = {{van Dishoeck}, Ewine F. and {Jonkheid}, Bastiaan and {van Hemert}, Marc C.},
        title = "{Photoprocesses in protoplanetary disks}",
      journal = {Faraday Discussions},
         year = 2006,
        month = jan,
       volume = {133},
        pages = {231},
          doi = {10.1039/b517564j},
       adsurl = {https://ui.adsabs.harvard.edu/abs/2006FaDi..133..231V}
}

@ARTICLE{arasa15,
       author = {{Arasa}, Carina and {Koning}, Jesper and {Kroes}, Geert-Jan and {Walsh}, Catherine and {van Dishoeck}, Ewine F.},
        title = "{Photodesorption of H$_{2}$O, HDO, and D$_{2}$O ice and its impact on fractionation}",
      journal = {\aap},
         year = 2015,
        month = mar,
       volume = {575},
          eid = {A121},
        pages = {A121},
          doi = {10.1051/0004-6361/201322695},
archivePrefix = {arXiv},
       eprint = {1503.00394},
 primaryClass = {astro-ph.GA},
       adsurl = {https://ui.adsabs.harvard.edu/abs/2015A&A...575A.121A}
}

@ARTICLE{komichi26,
       author = {{Komichi}, Yuto and {Aikawa}, Yuri and {Iwasaki}, Kazunari and {Furuya}, Kenji},
        title = "{Time-dependent chemical evolution during cloud formation: H$_{2}$-regulated chemistry in diffuse molecular cloud}",
      journal = {\mnras},
         year = 2026,
        month = feb,
       volume = {546},
       number = {2},
          eid = {stag027},
        pages = {stag027},
          doi = {10.1093/mnras/stag027},
archivePrefix = {arXiv},
       eprint = {2601.03441},
 primaryClass = {astro-ph.GA},
       adsurl = {https://ui.adsabs.harvard.edu/abs/2026MNRAS.546ag027K}
}

@ARTICLE{caselli02,
       author = {{Caselli}, Paola and {Stantcheva}, Tatiana and {Shalabiea}, Osama and {Shematovich}, Valery I. and {Herbst}, Eric},
        title = "{Deuterium fractionation on interstellar grains studied with modified rate equations and a Monte Carlo approach}",
      journal = {\planss},
         year = 2002,
        month = oct,
       volume = {50},
       number = {12-13},
        pages = {1257-1266},
          doi = {10.1016/S0032-0633(02)00092-2},
archivePrefix = {arXiv},
       eprint = {astro-ph/0202368},
 primaryClass = {astro-ph},
       adsurl = {https://ui.adsabs.harvard.edu/abs/2002P&SS...50.1257C}
}

@ARTICLE{cooke18,
       author = {{Cooke}, Ryan J. and {Pettini}, Max and {Steidel}, Charles C.},
        title = "{One Percent Determination of the Primordial Deuterium Abundance}",
      journal = {\apj},
         year = 2018,
        month = mar,
       volume = {855},
       number = {2},
          eid = {102},
        pages = {102},
          doi = {10.3847/1538-4357/aaab53},
archivePrefix = {arXiv},
       eprint = {1710.11129},
 primaryClass = {astro-ph.CO},
       adsurl = {https://ui.adsabs.harvard.edu/abs/2018ApJ...855..102C}
}

@ARTICLE{whittet13,
       author = {{Whittet}, D.~C.~B. and {Poteet}, C.~A. and {Chiar}, J.~E. and {Pagani}, L. and {Bajaj}, V.~M. and {Horne}, D. and {Shenoy}, S.~S. and {Adamson}, A.~J.},
        title = "{Ice and Dust in the Prestellar Dark Cloud Lynds 183: Preplanetary Matter at the Lowest Temperatures}",
      journal = {\apj},
         year = 2013,
        month = sep,
       volume = {774},
       number = {2},
          eid = {102},
        pages = {102},
          doi = {10.1088/0004-637X/774/2/102},
       adsurl = {https://ui.adsabs.harvard.edu/abs/2013ApJ...774..102W}
}

@ARTICLE{jorgensen18,
       author = {{J{\o}rgensen}, J.~K. and {M{\"u}ller}, H.~S.~P. and {Calcutt}, H. and {Coutens}, A. and {Drozdovskaya}, M.~N. and {{\"O}berg}, K.~I. and {Persson}, M.~V. and {Taquet}, V. and {van Dishoeck}, E.~F. and {Wampfler}, S.~F.},
        title = "{The ALMA-PILS survey: isotopic composition of oxygen-containing complex organic molecules toward IRAS 16293-2422B}",
      journal = {\aap},
         year = 2018,
        month = dec,
       volume = {620},
          eid = {A170},
        pages = {A170},
          doi = {10.1051/0004-6361/201731667},
archivePrefix = {arXiv},
       eprint = {1808.08753},
 primaryClass = {astro-ph.SR},
       adsurl = {https://ui.adsabs.harvard.edu/abs/2018A&A...620A.170J}
}

@INPROCEEDINGS{Ward-Thompson07,
       author = {{Ward-Thompson}, D. and {Andr{\'e}}, P. and {Crutcher}, R. and {Johnstone}, D. and {Onishi}, T. and {Wilson}, C.},
        title = "{An Observational Perspective of Low-Mass Dense Cores II: Evolution Toward the Initial Mass Function}",
    booktitle = {Protostars and Planets V},
         year = 2007,
       editor = {{Reipurth}, Bo and {Jewitt}, David and {Keil}, Klaus},
        month = jan,
        pages = {33},
          doi = {10.48550/arXiv.astro-ph/0603474},
archivePrefix = {arXiv},
       eprint = {astro-ph/0603474},
 primaryClass = {astro-ph},
       adsurl = {https://ui.adsabs.harvard.edu/abs/2007prpl.conf...33W}
}

@ARTICLE{takemura23,
       author = {{Takemura}, Hideaki and {Nakamura}, Fumitaka and {Arce}, H{\'e}ctor G. and {Schneider}, Nicola and {Ossenkopf-Okada}, Volker and {Kong}, Shuo and {Ishii}, Shun and {Dobashi}, Kazuhito and {Shimoikura}, Tomomi and {Sanhueza}, Patricio and {Tsukagoshi}, Takashi and {Padoan}, Paolo and {Klessen}, Ralf S. and {Goldsmith}, Paul. F. and {Burkhart}, Blakesley and {Lis}, Dariusz C. and {S{\'a}nchez-Monge}, {\'A}lvaro and {Shimajiri}, Yoshito and {Kawabe}, Ryohei},
        title = "{CARMA-NRO Orion Survey: Unbiased Survey of Dense Cores and Core Mass Functions in Orion A}",
      journal = {\apjs},
         year = 2023,
        month = feb,
       volume = {264},
       number = {2},
          eid = {35},
        pages = {35},
          doi = {10.3847/1538-4365/aca4d4},
archivePrefix = {arXiv},
       eprint = {2211.10215},
 primaryClass = {astro-ph.GA},
       adsurl = {https://ui.adsabs.harvard.edu/abs/2023ApJS..264...35T}
}

@ARTICLE{taylor25,
       author = {{Taylor}, Aster G. and {Seligman}, Darryl Z.},
        title = "{The Kinematic Age of 3I/ATLAS and Its Implications for Early Planet Formation}",
      journal = {\apjl},
         year = 2025,
        month = sep,
       volume = {990},
       number = {1},
          eid = {L14},
        pages = {L14},
          doi = {10.3847/2041-8213/adfa28},
archivePrefix = {arXiv},
       eprint = {2507.08111},
 primaryClass = {astro-ph.EP},
       adsurl = {https://ui.adsabs.harvard.edu/abs/2025ApJ...990L..14T}
}

@ARTICLE{hopkins25,
       author = {{Hopkins}, Matthew J. and {Dorsey}, Rosemary C. and {Forbes}, John C. and {Bannister}, Michele T. and {Lintott}, Chris J. and {Leicester}, Brayden},
        title = "{From a Different Star: 3I/ATLAS in the Context of the {\={O}}tautahi─Oxford Interstellar Object Population Model}",
      journal = {\apjl},
         year = 2025,
        month = sep,
       volume = {990},
       number = {2},
          eid = {L30},
        pages = {L30},
          doi = {10.3847/2041-8213/adfbf4},
archivePrefix = {arXiv},
       eprint = {2507.05318},
 primaryClass = {astro-ph.EP},
       adsurl = {https://ui.adsabs.harvard.edu/abs/2025ApJ...990L..30H}
}

@ARTICLE{lee15,
       author = {{Lee}, Jeong-Eun and {Bergin}, Edwin A.},
        title = "{The D/H Ratio of Water Ice at Low Temperatures}",
      journal = {\apj},
         year = 2015,
        month = jan,
       volume = {799},
       number = {1},
          eid = {104},
        pages = {104},
          doi = {10.1088/0004-637X/799/1/104},
archivePrefix = {arXiv},
       eprint = {1411.4231},
 primaryClass = {astro-ph.GA},
       adsurl = {https://ui.adsabs.harvard.edu/abs/2015ApJ...799..104L}
}

@ARTICLE{moon25,
       author = {{Moon}, Sanghyuk and {Ostriker}, Eve C.},
        title = "{Prestellar Cores in Turbulent Clouds: Observational Perspectives on Structure, Kinematics, and Lifetime}",
      journal = {\apj},
         year = 2025,
        month = nov,
       volume = {994},
       number = {1},
          eid = {79},
        pages = {79},
          doi = {10.3847/1538-4357/ae04eb},
archivePrefix = {arXiv},
       eprint = {2509.07083},
 primaryClass = {astro-ph.GA},
       adsurl = {https://ui.adsabs.harvard.edu/abs/2025ApJ...994...79M}
}

@ARTICLE{zeng25,
       author = {{Zeng}, Shaoshan and {Jeong}, Jae-Hong and {Oyama}, Takahiro and {Lee}, Jeong-Eun and {Yang}, Yao-Lun and {Sakai}, Nami},
        title = "{Determining the Methanol Deuteration in the Disk Around V883 Orionis with Laboratory Measured Spectroscopy}",
      journal = {\aj},
         year = 2025,
        month = jul,
       volume = {170},
       number = {1},
          eid = {33},
        pages = {33},
          doi = {10.3847/1538-3881/add733},
archivePrefix = {arXiv},
       eprint = {2506.07794},
 primaryClass = {astro-ph.SR},
       adsurl = {https://ui.adsabs.harvard.edu/abs/2025AJ....170...33Z}
}

@ARTICLE{hollenbach79,
       author = {{Hollenbach}, D. and {McKee}, C.~F.},
        title = "{Molecule formation and infrared emission in fast interstellar shocks. I. Physical processes.}",
      journal = {\apjs},
         year = 1979,
        month = nov,
       volume = {41},
        pages = {555-592},
          doi = {10.1086/190631},
       adsurl = {https://ui.adsabs.harvard.edu/abs/1979ApJS...41..555H}
}

@ARTICLE{furuya24,
       author = {{Furuya}, Kenji},
        title = "{A Framework for Incorporating Binding Energy Distribution in Gas-ice Astrochemical Models}",
      journal = {\apj},
         year = 2024,
        month = oct,
       volume = {974},
       number = {1},
          eid = {115},
        pages = {115},
          doi = {10.3847/1538-4357/ad6b20},
archivePrefix = {arXiv},
       eprint = {2408.02958},
 primaryClass = {astro-ph.GA},
       adsurl = {https://ui.adsabs.harvard.edu/abs/2024ApJ...974..115F}
}

@ARTICLE{pagani92,
       author = {{Pagani}, L. and {Salez}, M. and {Wannier}, P.~G.},
        title = "{The chemistry of H2D+ in cold clouds.}",
      journal = {\aap},
         year = 1992,
        month = may,
       volume = {258},
        pages = {479-488},
       adsurl = {https://ui.adsabs.harvard.edu/abs/1992A&A...258..479P}
}

@ARTICLE{linsky03,
       author = {{Linsky}, Jeffrey L.},
        title = "{Atomic Deuterium/Hydrogen in the Galaxy}",
      journal = {\ssr},
         year = 2003,
        month = apr,
       volume = {106},
       number = {1},
        pages = {49-60},
          doi = {10.1023/A:1024673217736},
archivePrefix = {arXiv},
       eprint = {astro-ph/0309099},
 primaryClass = {astro-ph},
       adsurl = {https://ui.adsabs.harvard.edu/abs/2003SSRv..106...49L}
}

@ARTICLE{habing68,
       author = {{Habing}, H.~J.},
        title = "{The interstellar radiation density between 912 A and 2400 A}",
      journal = {\bain},
         year = 1968,
        month = jan,
       volume = {19},
        pages = {421},
       adsurl = {https://ui.adsabs.harvard.edu/abs/1968BAN....19..421H}
}

@ARTICLE{mandt24,
       author = {{Mandt}, Kathleen E. and {Lustig-Yaeger}, Jacob and {Luspay-Kuti}, Adrienn and {Wurz}, Peter and {Bodewits}, Dennis and {Fuselier}, Stephen A. and {Mousis}, Olivier and {Petrinec}, Steven M. and {Trattner}, Karlheinz J.},
        title = "{A nearly terrestrial D/H for comet 67P/Churyumov-Gerasimenko}",
      journal = {Science Advances},
         year = 2024,
        month = nov,
       volume = {10},
       number = {46},
          eid = {eadp2191},
        pages = {eadp2191},
          doi = {10.1126/sciadv.adp2191},
       adsurl = {https://ui.adsabs.harvard.edu/abs/2024SciA...10P2191M}
}

@ARTICLE{honvault12,
       author = {{Honvault}, P. and {Jorfi}, M. and {Gonz{\'a}lez-Lezana}, T. and {Faure}, A. and {Pagani}, L.},
        title = "{Erratum: Otho-Para H$_{2}$ Conversion by Proton Exchange at Low Temperature: An Accurate Quantum Mechanical Study [Phys. Rev. Lett. 107, 023201 (2011)]}",
      journal = {\prl},
         year = 2012,
        month = mar,
       volume = {108},
       number = {10},
          eid = {109903},
        pages = {109903},
          doi = {10.1103/PhysRevLett.108.109903},
       adsurl = {https://ui.adsabs.harvard.edu/abs/2012PhRvL.108j9903H}
}

@ARTICLE{coutens14,
       author = {{Coutens}, A. and {Vastel}, C. and {Hincelin}, U. and {Herbst}, E. and {Lis}, D.~C. and {Chavarr{\'\i}a}, L. and {G{\'e}rin}, M. and {van der Tak}, F.~F.~S. and {Persson}, C.~M. and {Goldsmith}, P.~F. and {Caux}, E.},
        title = "{Water deuterium fractionation in the high-mass star-forming region G34.26+0.15 based on Herschel/HIFI data}",
      journal = {\mnras},
         year = 2014,
        month = dec,
       volume = {445},
       number = {2},
        pages = {1299-1313},
          doi = {10.1093/mnras/stu1816},
archivePrefix = {arXiv},
       eprint = {1409.1092},
 primaryClass = {astro-ph.SR},
       adsurl = {https://ui.adsabs.harvard.edu/abs/2014MNRAS.445.1299C}
}

@ARTICLE{aikawa12,
       author = {{Aikawa}, Y. and {Wakelam}, V. and {Hersant}, F. and {Garrod}, R.~T. and {Herbst}, E.},
        title = "{From Prestellar to Protostellar Cores. II. Time Dependence and Deuterium Fractionation}",
      journal = {\apj},
         year = 2012,
        month = nov,
       volume = {760},
       number = {1},
          eid = {40},
        pages = {40},
          doi = {10.1088/0004-637X/760/1/40},
archivePrefix = {arXiv},
       eprint = {1210.2476},
 primaryClass = {astro-ph.GA},
       adsurl = {https://ui.adsabs.harvard.edu/abs/2012ApJ...760...40A}
}

@ARTICLE{watson76,
       author = {{Watson}, William D.},
        title = "{Interstellar molecule reactions}",
      journal = {Reviews of Modern Physics},
         year = 1976,
        month = oct,
       volume = {48},
       number = {4},
        pages = {513-552},
          doi = {10.1103/RevModPhys.48.513},
       adsurl = {https://ui.adsabs.harvard.edu/abs/1976RvMP...48..513W}
}

@ARTICLE{gerlich02,
       author = {{Gerlich}, Dieter and {Herbst}, Eric and {Roueff}, Evelyne},
        title = "{H$_{3}$$^{+}$+HD<-- >H$_{2}$D$^{+}$+H$_{2}$: low-temperature laboratory measurements and interstellar implications}",
      journal = {\planss},
         year = 2002,
        month = oct,
       volume = {50},
       number = {12-13},
        pages = {1275-1285},
          doi = {10.1016/S0032-0633(02)00094-6},
       adsurl = {https://ui.adsabs.harvard.edu/abs/2002P&SS...50.1275G}
}

@ARTICLE{honvault11,
       author = {{Honvault}, P. and {Jorfi}, M. and {Gonz{\'a}lez-Lezana}, T. and {Faure}, A. and {Pagani}, L.},
        title = "{Ortho-Para H$_{2}$ Conversion by Proton Exchange at Low Temperature: An Accurate Quantum Mechanical Study}",
      journal = {\prl},
         year = 2011,
        month = jul,
       volume = {107},
       number = {2},
          eid = {023201},
        pages = {023201},
          doi = {10.1103/PhysRevLett.107.023201},
       adsurl = {https://ui.adsabs.harvard.edu/abs/2011PhRvL.107b3201H}
}

@ARTICLE{watanabe10,
       author = {{Watanabe}, Naoki and {Kimura}, Yuki and {Kouchi}, Akira and {Chigai}, Takeshi and {Hama}, Tetsuya and {Pirronello}, Valerio},
        title = "{Direct Measurements of Hydrogen Atom Diffusion and the Spin Temperature of Nascent H$_{2}$ Molecule on Amorphous Solid Water}",
      journal = {\apjl},
         year = 2010,
        month = may,
       volume = {714},
       number = {2},
        pages = {L233-L237},
          doi = {10.1088/2041-8205/714/2/L233},
       adsurl = {https://ui.adsabs.harvard.edu/abs/2010ApJ...714L.233W}
}

@ARTICLE{goldsmith05,
       author = {{Goldsmith}, P.~F. and {Li}, D.},
        title = "{H I Narrow Self-Absorption in Dark Clouds: Correlations with Molecular Gas and Implications for Cloud Evolution and Star Formation}",
      journal = {\apj},
         year = 2005,
        month = apr,
       volume = {622},
       number = {2},
        pages = {938-958},
          doi = {10.1086/428032},
archivePrefix = {arXiv},
       eprint = {astro-ph/0412427},
 primaryClass = {astro-ph},
       adsurl = {https://ui.adsabs.harvard.edu/abs/2005ApJ...622..938G}
}

@BOOK{tielens05,
       author = {{Tielens}, A.~G.~G.~M.},
        title = "{The Physics and Chemistry of the Interstellar Medium}",
         year = 2005,
       adsurl = {https://ui.adsabs.harvard.edu/abs/2005pcim.book.....T}
}

@INPROCEEDINGS{nomura23,
       author = {{Nomura}, H. and {Furuya}, K. and {Cordiner}, M.~A. and {Charnley}, S.~B. and {Alexander}, C.~M. O'D. and {Nixon}, C.~A. and {Guzman}, V.~V. and {Yurimoto}, H. and {Tsukagoshi}, T. and {Iino}, T.},
        title = "{The Isotopic Links from Planet Forming Regions to the Solar System}",
    booktitle = {Protostars and Planets VII},
         year = 2023,
       editor = {{Inutsuka}, S. and {Aikawa}, Y. and {Muto}, T. and {Tomida}, K. and {Tamura}, M.},
       series = {Astronomical Society of the Pacific Conference Series},
       volume = {534},
        month = jul,
        pages = {1075},
       adsurl = {https://ui.adsabs.harvard.edu/abs/2023ASPC..534.1075N}
}

@ARTICLE{furuya17,
       author = {{Furuya}, K. and {Drozdovskaya}, M.~N. and {Visser}, R. and {van Dishoeck}, E.~F. and {Walsh}, C. and {Harsono}, D. and {Hincelin}, U. and {Taquet}, V.},
        title = "{Water delivery from cores to disks: Deuteration as a probe of the prestellar inheritance of H$_{2}$O}",
      journal = {\aap},
         year = 2017,
        month = mar,
       volume = {599},
          eid = {A40},
        pages = {A40},
          doi = {10.1051/0004-6361/201629269},
archivePrefix = {arXiv},
       eprint = {1610.07286},
 primaryClass = {astro-ph.GA},
       adsurl = {https://ui.adsabs.harvard.edu/abs/2017A&A...599A..40F}
}

@ARTICLE{oba12,
       author = {{Oba}, Y. and {Watanabe}, N. and {Hama}, T. and {Kuwahata}, K. and {Hidaka}, H. and {Kouchi}, A.},
        title = "{Water Formation through a Quantum Tunneling Surface Reaction, OH + H$_{2}$, at 10 K}",
      journal = {\apj},
         year = 2012,
        month = apr,
       volume = {749},
       number = {1},
          eid = {67},
        pages = {67},
          doi = {10.1088/0004-637X/749/1/67},
archivePrefix = {arXiv},
       eprint = {1202.1035},
 primaryClass = {astro-ph.GA},
       adsurl = {https://ui.adsabs.harvard.edu/abs/2012ApJ...749...67O}
}

@ARTICLE{boogert15,
       author = {{Boogert}, A.~C. Adwin and {Gerakines}, Perry A. and {Whittet}, Douglas C.~B.},
        title = "{Observations of the icy universe.}",
      journal = {\araa},
         year = 2015,
        month = aug,
       volume = {53},
        pages = {541-581},
          doi = {10.1146/annurev-astro-082214-122348},
archivePrefix = {arXiv},
       eprint = {1501.05317},
 primaryClass = {astro-ph.GA},
       adsurl = {https://ui.adsabs.harvard.edu/abs/2015ARA&A..53..541B}
}

@ARTICLE{aikawa99,
       author = {{Aikawa}, Yuri and {Herbst}, Eric},
        title = "{Deuterium Fractionation in Protoplanetary Disks}",
      journal = {\apj},
         year = 1999,
        month = nov,
       volume = {526},
       number = {1},
        pages = {314-326},
          doi = {10.1086/307973},
       adsurl = {https://ui.adsabs.harvard.edu/abs/1999ApJ...526..314A}
}

@ARTICLE{hasegawa93,
       author = {{Hasegawa}, T.~I. and {Herbst}, E.},
        title = "{Three-Phase Chemical Models of Dense Interstellar Clouds - Gas Dust Particle Mantles and Dust Particle Surfaces}",
      journal = {\mnras},
         year = 1993,
        month = aug,
       volume = {263},
        pages = {589},
          doi = {10.1093/mnras/263.3.589},
       adsurl = {https://ui.adsabs.harvard.edu/abs/1993MNRAS.263..589H}
}

@Article{matplotlib,
  Author    = {Hunter, J. D.},
  Title     = {Matplotlib: A 2D graphics environment},
  Journal   = {Computing in Science \& Engineering},
  Volume    = {9},
  Number    = {3},
  Pages     = {90--95},
  publisher = {IEEE COMPUTER SOC},
  doi       = {10.1109/MCSE.2007.55},
  year      = 2007
}

@article{yang13,
title = {The D/H ratio of water in the solar nebula during its formation and evolution},
journal = {Icarus},
volume = {226},
number = {1},
pages = {256-267},
year = {2013},
issn = {0019-1035},
doi = {https://doi.org/10.1016/j.icarus.2013.05.027},
url = {https://www.sciencedirect.com/science/article/pii/S0019103513002273},
author = {Le Yang and Fred J. Ciesla and Conel M.O’D. Alexander}
}

@ARTICLE{owen15,
       author = {{Owen}, James E. and {Jacquet}, Emmanuel},
        title = "{Astro- and cosmochemical consequences of accretion bursts - I. The D/H ratio of water}",
      journal = {\mnras},
         year = 2015,
        month = feb,
       volume = {446},
       number = {4},
        pages = {3285-3296},
          doi = {10.1093/mnras/stu2254},
archivePrefix = {arXiv},
       eprint = {1410.6844},
 primaryClass = {astro-ph.SR},
       adsurl = {https://ui.adsabs.harvard.edu/abs/2015MNRAS.446.3285O}
}

@ARTICLE{remy-ruyer14,
       author = {{R{\'e}my-Ruyer}, A. and {Madden}, S.~C. and {Galliano}, F. and {Galametz}, M. and {Takeuchi}, T.~T. and {Asano}, R.~S. and {Zhukovska}, S. and {Lebouteiller}, V. and {Cormier}, D. and {Jones}, A. and {Bocchio}, M. and {Baes}, M. and {Bendo}, G.~J. and {Boquien}, M. and {Boselli}, A. and {DeLooze}, I. and {Doublier-Pritchard}, V. and {Hughes}, T. and {Karczewski}, O. {\L}. and {Spinoglio}, L.},
        title = "{Gas-to-dust mass ratios in local galaxies over a 2 dex metallicity range}",
      journal = {\aap},
         year = 2014,
        month = mar,
       volume = {563},
          eid = {A31},
        pages = {A31},
          doi = {10.1051/0004-6361/201322803},
archivePrefix = {arXiv},
       eprint = {1312.3442},
 primaryClass = {astro-ph.GA},
       adsurl = {https://ui.adsabs.harvard.edu/abs/2014A&A...563A..31R}
}

@ARTICLE{galliano18,
       author = {{Galliano}, Fr{\'e}d{\'e}ric and {Galametz}, Maud and {Jones}, Anthony P.},
        title = "{The Interstellar Dust Properties of Nearby Galaxies}",
      journal = {\araa},
         year = 2018,
        month = sep,
       volume = {56},
        pages = {673-713},
          doi = {10.1146/annurev-astro-081817-051900},
archivePrefix = {arXiv},
       eprint = {1711.07434},
 primaryClass = {astro-ph.GA},
       adsurl = {https://ui.adsabs.harvard.edu/abs/2018ARA&A..56..673G}
}

@ARTICLE{relano22,
       author = {{Rela{\~n}o}, M. and {De Looze}, I. and {Saintonge}, A. and {Hou}, K.-C. and {Romano}, L.~E.~C. and {Nagamine}, K. and {Hirashita}, H. and {Aoyama}, S. and {Lamperti}, I. and {Lisenfeld}, U. and {Smith}, M.~W.~L. and {Chastenet}, J. and {Xiao}, T. and {Gao}, Y. and {Sargent}, M. and {van der Giessen}, S.~A.},
        title = "{Dust grain size evolution in local galaxies: a comparison between observations and simulations}",
      journal = {\mnras},
         year = 2022,
        month = oct,
       volume = {515},
       number = {4},
        pages = {5306-5334},
          doi = {10.1093/mnras/stac2108},
archivePrefix = {arXiv},
       eprint = {2207.13196},
 primaryClass = {astro-ph.GA},
       adsurl = {https://ui.adsabs.harvard.edu/abs/2022MNRAS.515.5306R}
}

@ARTICLE{giannetti17,
       author = {{Giannetti}, A. and {Leurini}, S. and {K{\"o}nig}, C. and {Urquhart}, J.~S. and {Pillai}, T. and {Brand}, J. and {Kauffmann}, J. and {Wyrowski}, F. and {Menten}, K.~M.},
        title = "{Galactocentric variation of the gas-to-dust ratio and its relation with metallicity}",
      journal = {\aap},
         year = 2017,
        month = oct,
       volume = {606},
          eid = {L12},
        pages = {L12},
          doi = {10.1051/0004-6361/201731728},
archivePrefix = {arXiv},
       eprint = {1710.05721},
 primaryClass = {astro-ph.GA},
       adsurl = {https://ui.adsabs.harvard.edu/abs/2017A&A...606L..12G}
}

@ARTICLE{cleeves14,
       author = {{Cleeves}, L. Ilsedore and {Bergin}, Edwin A. and {Alexander}, Conel M.~O. 'D. and {Du}, Fujun and {Graninger}, Dawn and {{\"O}berg}, Karin I. and {Harries}, Tim J.},
        title = "{The ancient heritage of water ice in the solar system}",
      journal = {Science},
         year = 2014,
        month = sep,
       volume = {345},
       number = {6204},
        pages = {1590-1593},
          doi = {10.1126/science.1258055},
archivePrefix = {arXiv},
       eprint = {1409.7398},
 primaryClass = {astro-ph.SR},
       adsurl = {https://ui.adsabs.harvard.edu/abs/2014Sci...345.1590C}
}

@ARTICLE{hocuk17,
       author = {{Hocuk}, S. and {Sz{\H{u}}cs}, L. and {Caselli}, P. and {Cazaux}, S. and {Spaans}, M. and {Esplugues}, G.~B.},
        title = "{Parameterizing the interstellar dust temperature}",
      journal = {\aap},
         year = 2017,
        month = aug,
       volume = {604},
          eid = {A58},
        pages = {A58},
          doi = {10.1051/0004-6361/201629944},
archivePrefix = {arXiv},
       eprint = {1704.02763},
 primaryClass = {astro-ph.GA},
       adsurl = {https://ui.adsabs.harvard.edu/abs/2017A&A...604A..58H}
}

@ARTICLE{cuppen07,
       author = {{Cuppen}, H.~M. and {Herbst}, Eric},
        title = "{Simulation of the Formation and Morphology of Ice Mantles on Interstellar Grains}",
      journal = {\apj},
         year = 2007,
        month = oct,
       volume = {668},
       number = {1},
        pages = {294-309},
          doi = {10.1086/521014},
archivePrefix = {arXiv},
       eprint = {0707.2744},
 primaryClass = {astro-ph},
       adsurl = {https://ui.adsabs.harvard.edu/abs/2007ApJ...668..294C}
}

@ARTICLE{furuya15,
       author = {{Furuya}, K. and {Aikawa}, Y. and {Hincelin}, U. and {Hassel}, G.~E. and {Bergin}, E.~A. and {Vasyunin}, A.~I. and {Herbst}, E.},
        title = "{Water deuteration and ortho-to-para nuclear spin ratio of H$_{2}$ in molecular clouds formed via the accumulation of H I gas}",
      journal = {\aap},
         year = 2015,
        month = dec,
       volume = {584},
          eid = {A124},
        pages = {A124},
          doi = {10.1051/0004-6361/201527050},
archivePrefix = {arXiv},
       eprint = {1510.05135},
 primaryClass = {astro-ph.GA},
       adsurl = {https://ui.adsabs.harvard.edu/abs/2015A&A...584A.124F}
}

@ARTICLE{furuya16,
       author = {{Furuya}, K. and {van Dishoeck}, E.~F. and {Aikawa}, Y.},
        title = "{Reconstructing the history of water ice formation from HDO/H$_{2}$O and D$_{2}$O/HDO ratios in protostellar cores}",
      journal = {\aap},
         year = 2016,
        month = feb,
       volume = {586},
          eid = {A127},
        pages = {A127},
          doi = {10.1051/0004-6361/201527579},
archivePrefix = {arXiv},
       eprint = {1512.04291},
 primaryClass = {astro-ph.GA},
       adsurl = {https://ui.adsabs.harvard.edu/abs/2016A&A...586A.127F}
}

@ARTICLE{furuya19,
       author = {{Furuya}, Kenji and {Aikawa}, Yuri and {Hama}, Tetsuya and {Watanabe}, Naoki},
        title = "{H$_{2}$ Ortho-Para Spin Conversion on Inhomogeneous Grain Surfaces}",
      journal = {\apj},
         year = 2019,
        month = sep,
       volume = {882},
       number = {2},
          eid = {172},
        pages = {172},
          doi = {10.3847/1538-4357/ab3790},
archivePrefix = {arXiv},
       eprint = {1908.01966},
 primaryClass = {astro-ph.GA},
       adsurl = {https://ui.adsabs.harvard.edu/abs/2019ApJ...882..172F}
}

@ARTICLE{draine78,
       author = {{Draine}, B.~T.},
        title = "{Photoelectric heating of interstellar gas.}",
      journal = {\apjs},
         year = 1978,
        month = apr,
       volume = {36},
        pages = {595-619},
          doi = {10.1086/190513},
       adsurl = {https://ui.adsabs.harvard.edu/abs/1978ApJS...36..595D}
}
\bibliographystyle{aasjournal}



\end{document}